\documentclass[twocolumn,amsmath,amssymb,floatfix,superscriptaddress]{revtex4}
\usepackage[dvips]{graphicx}
\usepackage{amsfonts}
\usepackage{dcolumn}
\usepackage{bm}
\usepackage{color}
\usepackage{epsfig}
\usepackage{hyperref}
\usepackage{bbold}
\usepackage[normalem]{ulem}

\newcommand{\be}{\begin{equation}}
\newcommand{\ee}{\end{equation}}
\newcommand{\bea}{\begin{eqnarray}}
\newcommand{\eea}{\end{eqnarray}}

\def\a{\alpha}

\def\G{\Gamma}
\def\d{\delta}
\def\D{\Delta}
\def\e{\epsilon}

\def\th{\theta}

\def\L{\Lambda}
\def\m{\mu}
\def\n{\nu}

\def\p{\pi}

\def\r{\rho}
\def\s{\sigma}
\def\S{\Sigma}
\def\t{\tau}

\def\vf{\varphi}
\def\F{\Phi}

\def\w{\omega}
\def\W{\Omega}

\def\Q{\Psi}

\def\bgs{\mbox{\boldmath $\sigma$}}

\def\bld{{\mathbf d}}
\def\ble{{\mathbf e}}

\def\blj{{\mathbf j}}

\def\blr{{\mathbf r}}
\def\bls{{\mathbf s}}

\def\blx{{\mathbf x}}

\def\blE{{\mathbf E}}

\def\callA{\mbox{$\mathcal{A}$}}

\def\callC{\mbox{$\mathcal{C}$}}

\def\callG{\mbox{$\mathcal{G}$}}

\def\callI{\mbox{$\mathcal{I}$}}
\def\callJ{\mbox{$\mathcal{J}$}}

\def\callS{\mbox{$\mathcal{S}$}}

\def\iif{\infty}
\def\bra{\langle}
\def\ket{\rangle}
\def\grad{\mbox{\boldmath $\nabla$}}
\def\Tr{{\rm Tr}}
\def\Re{{\rm Re}}
\def\Im{{\rm Im}}

\def\1op{\hat{\mathbbm{1}}}
\def\nn{\nonumber}

\begin{document}

\title{CHEERS: A tool for Correlated Hole-Electron Evolution from 
Real-time Simulations}

\author{E. Perfetto}
\affiliation{CNR-ISM, Division of Ultrafast Processes in Materials 
(FLASHit), Area della ricerca di Roma 1, Monterotondo Scalo, Italy}
\affiliation{Dipartimento di Fisica, 
Universit\`a di Roma Tor Vergata, Via della Ricerca Scientifica, 00133 Rome, Italy}

\author{G. Stefanucci}
\affiliation{Dipartimento di Fisica, Universit\`a di Roma Tor Vergata, Via della Ricerca Scientifica, 00133 Rome, Italy}
\affiliation{INFN, Sezione di Roma Tor Vergata, Via della Ricerca Scientifica 1, 00133 Roma, Italy}
\date{\today}

\begin{abstract}
We put forward a practical nonequilibrium Green's 
function (NEGF) scheme to perform real-time evolutions of many-body 
interacting systems driven out of equilibrium by external fields. 
CHEERS is a computational tool to solve the NEGF equation of motion 
in the so called generalized Kadanoff-Baym ansatz and it can 
be used for model systems as well as first-principles Hamiltonians.
Dynamical correlation (or memory) effects are added to the Hartree-Fock 
dynamics through a many-body self-energy. Applications to
time-dependent quantum transport, time-resolved photoabsorption and 
other ultrafast phenomena are discussed.
\end{abstract}
\maketitle

\section{Introduction}

Although the laws of quantum mechanics have been formulated almost a 
century ago, the behavior of quantum matter under (extreme) 
nonequilibrium conditions remains still largely unexplored.
Modern advances in laser technology~\cite{RevModPhys.81.163,Gallmann-review,Nisoli-review} 
make today possible to film, 
with an unprecedented time resolution,
the genesis and development of  
photoemission processes, exciton formation, charge transfers, charge 
migrations, Auger 
decays and other ultrafast 
phenomena. This is the realm of attosecond physics which calls for  
accurate theories and efficient numerical 
schemes to predict the evolution of  many-body quantum systems.

One of the most versatile formalism to deal with the quantum many-body 
problem 
is the diagrammatic  Green's function theory. The extension to out-of-equilibrium 
situations is known as the Non-Equilibrium Green's Function (NEGF) 
theory~\cite{danielewicz1984quantum,svl-book,balzer2012nonequilibrium} 
and the fundamental equations, known as Kadanoff-Baym equations (KBE), 
date back to the mid 
sixties~\cite{konstantinov1961diagram,kadanoff1962quantum,keldysh1965diagram}. 
Despite the enormous advance in computational capabilities the KBE 
are still rather burdensome to solve numerically. In fact, their implementations 
have been so far restricted to atoms, diatomic molecules or 
model 
systems~\cite{PhysRevLett.84.1768,PhysRevLett.98.153004,PhysRevB.79.245306,mssvl.2008,mssvl.2009,pva.2009,pva.2010,PhysRevA.81.022510,PhysRevA.82.033427,PhysRevB.93.054303}. 
In the mid-1980s Lipavsky {\em et al.}~\cite{PhysRevB.34.6933} 
proposed the so called Generalized Kadanoff-Baym Ansatz (GKBA) to 
collapse the KBE for the two-times Green's function into a single equation 
for the one-time one-particle density matrix, thus drastically 
reducing the computational cost. The GKBA is
exact in the Hartree-Fock (HF) approximation and it is expected to be accurate
when the average time between two consecutive collisions is 
longer than the quasiparticle decay time (see Ref.~\cite{LPUvLS.2014}
for a recent discussion). 

The appealing feature of the GKBA is that the NEGF formalism is converted into a 
time-dependent 
density-matrix functional 
theory~\cite{PhysRevA.75.012506,Giesbertz2010,Giesbertz2014,Brics2016,Brics2017,Lackner-PRA2015,Lackner-JOPCS2015}
which shares a fundamental property with many-body perturbation 
theory, i.e., the systematic inclusion  of correlations through a proper selection 
of self-energy diagrams. Recent applications of the NEGF+GKBA 
approach include the nonequilibrium 
dynamics~\cite{HermannsPRB2014,CTPPBonitz2016} and many-body 
localization~\cite{BarLevReichman} of Hubbard 
clusters,
equilibrium absorption of sodium clusters~\cite{Pal2011},
transient absorption~\cite{PSMS.2015,Sangalli-2016,Pogna.2016} and  
carrier dynamics~\cite{SangalliEPL2015,PSMS.2016} of semiconductors.

In this work we describe CHEERS, a 
first-principles numerical tool based on NEGF+GKBA to simulate the time evolution of 
interacting  systems. CHEERS time-evolutions contain 
dynamical correlation (or memory) effects 
responsible for double (and multiple) excitations, decoherence-induced charge 
separation, Auger decays, shake-up dynamics, image-charge 
renormalizations, etc. Standard 
time-dependent HF simulations are recovered by simply 
switching off the effects of correlations. CHEERS has been 
already used to study the charge dynamics of model molecular 
junctions~\cite{LPUvLS.2014}, the 
transient photoabsorption spectrum of noble gas 
atoms~\cite{PUvLS.2015}, the formation of charge-transfer excitons 
and their subsequent  
separation in 
donor-acceptor complexes~\cite{C60paper2018}, the 
attosecond pulse-induced  charge migration in 
the phenylalanine aminoacid~\cite{PSMS.2018} and  time-resolved 
Auger decays~\cite{Covito2018}.

The paper is organized as follows. In Section~\ref{physicalsystemsec} 
we discuss the  physical systems of interest and write 
down the many-body Hamiltonian to describe them. The theoretical 
framework based on NEGF and GKBA is outlined in 
Section~\ref{theorysec} along with the NEGF equation of motion 
solved by CHEERS. In Section~\ref{cheerspp-sec} we list the  
observable quantities accessible from the solution of the NEGF 
equation. A description of the implementation details is 
given in Section~\ref{imp-det-sec}. Conclusions and outlooks are 
drawn in Section~\ref{conclusionssec}.

\section{Physical systems}
\label{physicalsystemsec}

We consider a quantum system with a finite number of nuclei. 
For the time being we fix the nuclear coordinates and 
focus on the electronic degrees of freedom. Let $\{\vf_{i}(\blx)\}$ be a set 
of localized orthonormal spin-orbitals suited to describe the 
equilibrium and nonequilibrium properties of the bound electrons. 
In our notation $\blx=\blr\s$ comprises a spatial coordinate $\blr$ 
and a spin-projection $\s$, and $\vf_{i}(\blx)=\bra i|\blx\ket$. The localized 
states could be generated by orthonormalizing a set of 
Slater-type orbitals (STO) or Gaussian-type 
orbitals (GTO) centered around each nucleus or they could be  
the bound Hartree-Fock (HF) orbitals or the bound Kohn-Sham (KS) orbitals 
resulting from some self-consistent HF or KS calculation 
respectively. We will give more details on the possible choice of the 
basis set in Section~\ref{imp-det-sec}. For each 
spin-orbital $\vf_{i}$ we define the corresponding annihilation and 
creation operator $\hat{c}_{i}$ and $\hat{c}^{\dag}_{i}$. The 
equilibrium Hamiltonian of the system in second quantization then reads
\be
\hat{H}^{\rm eq}_{\rm sys}=\sum_{ij}h^{\rm eq}_{ij}\hat{c}^{\dag}_{i}\hat{c}_{j}
+\frac{1}{2}\sum_{ijmn}v_{ijmn}
\hat{c}^{\dag}_{i}\hat{c}^{\dag}_{j}\hat{c}_{m}\hat{c}_{n}.
\label{eqsysH}
\ee
Here $h^{\rm eq}_{ij}$ are the matrix elements of the single-particle 
Hamiltonian (atomic units are used throughout): 
\be
h^{\rm eq}_{ij}\equiv\bra i|\,\frac{\hat{p}^{2}}{2}+\hat{V}_{\rm n}+\hat{V}_{\rm 
SO}|j\ket,
\ee
with $\hat{V}_{\rm n}$ the nuclear potential and $\hat{V}_{\rm SO}$ the 
spin-orbit interaction potential. The second term in 
Eq.~(\ref{eqsysH}) describes the electron-electron interaction with 
Coulomb integrals
\be
v_{ijmn}\equiv \int d\blx d\blx'\,
\frac{\vf_{i}^{\ast}(\blx)\vf_{j}^{\ast}(\blx')
\vf_{m}(\blx')\vf_{n}(\blx)}{|\blr-\blr'|},
\label{coulomb-int}
\ee
where $\int d\blx=\int d\blr\sum_{\s}$.

We are interested in studying the quantum evolution induced by 
an external electromagnetic field  
with spatial variations on length-scales much longer than the 
linear dimension of the system.  For 
nanometer-sized molecules this condition implies photon energies up to
1 keV and hence the possibility of photoionization. To describe 
photoionization processes it is  
necessary to extend the localized basis and include 
delocalized spin-orbitals $\{\vf_{\m}(\blx)\}$ for electrons in the 
continuum. Without loss of generality,
we choose the $\vf_{\m}$'s as eigenstates of the 
free-particle Hamiltonian far away from the system boundaries and we 
denote by $\e_{\m}$ their energy. 
We also require that 
the $\vf_{\m}$'s  are orthogonal to the $\vf_{i}$'s and
orthonormal between themeselves, i.e., $\bra\m|\m'\ket=\d_{\m\m'}$. 
An example of how to construct the $\vf_{\m}$'s is given in Section~\ref{imp-det-sec}.
We discard the Coulomb interaction between two electrons in the 
continuum since for weak pulses double ionization is 
strongly suppressed. Of course, if X-rays are used then a second (Auger) electron 
can be ejected. As we discuss in Section~\ref{theorysec}, 
in this case the approximation 
is justified provided that the photoelectron and the Auger electron have different 
energies.
The continuum Hamiltonian then reads
\be
\hat{H}_{\rm cont}=\sum_{\m}\e_{\m}\hat{c}^{\dag}_{\m}\hat{c}_{\m}\,.
\ee
Taking into account that the electric field $\blE$ is uniform  
for all localized states, we can write the interaction Hamiltonian between 
light and matter as
\be
\hat{H}^{\blE}(t)=\hat{H}^{\blE}_{\rm sys}(t)+\hat{H}^{\blE}_{\rm 
ion}(t),
\label{E-ham}
\ee
where 
\be
\hat{H}^{\blE}_{\rm sys}(t)
=\blE(t)\cdot
\sum_{ij}\bld_{ij}\hat{c}^{\dag}_{i}\hat{c}_{j}
\label{sysE-ham}
\ee
is the part responsible for reshuffling the electrons between  
localized states whereas
\be
\hat{H}^{\blE}_{\rm ion}(t)=
\blE(t)\cdot
\sum_{i\m}\left(\bld_{i\m}\hat{c}^{\dag}_{i}\hat{c}_{\m}+{\rm 
h.c.}\right)
\label{ionE-ham}
\ee
is the part responsible for photoionization processes. In 
Eqs.~(\ref{sysE-ham}) and (\ref{ionE-ham}) the vector of matrices 
$\bld$ is the dipole moment defined as 
\be
\bld_{ab}=\int d\blx \,
\vf_{a}^{\ast}(\blx)\,\blr\,\vf_{b}(\blx),
\ee
where $a$ and $b$ are indices 
either in the localized $\{i\}$ or delocalized $\{\m\}$ sector. In 
Eq.~(\ref{E-ham}) we are discarding transitions $\m\to\m'$ between delocalized 
electrons since we are mainly concerned with 
ultrafast fields (by the time the population of a $\m$-state 
becomes relevant the electric field vanishes).

During the first few femtoseconds after ionization from a (semi) core state, 
the  Auger decay is
one of the most relevant recombination channels. To account for the Auger 
effect in our description (and hence to deal with soft X-ray pulses 
too) we include the Coulomb matrix elements responsible for 
two localized (valence) electrons to scatter in one localized (core) 
electron and one delocalized (continuum) electron. In second 
quantization the Auger interaction Hamiltonian reads
\be
\hat{H}_{\rm Auger}=\sum_{ijm}\sum_{\m}v^{A}_{ijm\m}
\left(\hat{c}^{\dag}_{i}\hat{c}^{\dag}_{j}\hat{c}_{m}\hat{c}_{\m}
+{\rm h.c.}\right)\,,
\label{Auger-ham}
\ee
where the Coulomb integrals $v^{A}_{ijm\m}$ are defined as in 
Eq.~(\ref{coulomb-int}) with $\vf_{n}\to\vf_{\m}$.

In CHEERS the quantum system can also be contacted to 
metallic leads with which to exchange electrons and energy.
Arbitrary time-dependent voltages can be applied to the leads to 
study transient currents, steady-states, AC responses or other kind of 
transport properties. It is also possible to switch on a 
thermomechanical field~\cite{Luttinger-thermal} 
to calculate time-dependent thermal 
currents~\cite{EichPrincipiVentraVignale:14,EDVV.2016,CETSR.2018}. 
The combination of applied voltages and 
external laser pulses $\blE(t)$ can instead be used to access the optical properties of 
current-carrying molecular junctions~\cite{GalperinCSR2017}.
In all cases the leads are treated as 
noninteracting semi-infinite crystals with a finite cross section and described in terms of 
semi-infinite Bloch states $\vf_{\a k}(\blx)$, where $\a$ is the lead 
index and $k$ specifies the energy $\e_{\a k}$ of the Bloch state. 
The Bloch states have to be orthogonal to both the $\vf_{i}$'s and  
the $\vf_{\m}$'s. Thus, in a quantum transport setup the $\m$-states are the 
free-particle continuum states of the quantum system contacted to 
leads.
The second-quantized form of the leads Hamiltonian is
\be
\hat{H}_{\rm lead}(t)=\sum_{\a}\sum_{k}
\F_{\a}(t)\left(\e_{\a k}+V_{\a}(t)\right)
\hat{c}^{\dag}_{\a k}\hat{c}_{\a k}\,,
\ee
where $V_{\a}(t)$ is the applied voltage and $\F_{\a}(t)=T_{\a}(t)/T$ 
is the ratio between the temperature at time $t$ and the equilibrium 
temperature (the thermomechanical field is therefore 
$\Q_{\a}=\F_{\a}-1$).
The contact Hamiltonian responsible for electron tunneling has the 
general form
\be
\hat{H}_{\rm tun}=\sum_{i,\a k}
\left(T_{i,\a k}\,\hat{c}^{\dag}_{i}\hat{c}_{\a k}+{\rm h.c.}\right),
\ee
where $T_{i,\a k}$ is the tunneling amplitude, i.e.,
the matrix element of the single-particle 
Hamiltonian, between the states $\vf_{i}$ and $\vf_{\a k}$.

\begin{figure}[t]
\centerline{
\includegraphics[width=0.49\textwidth]{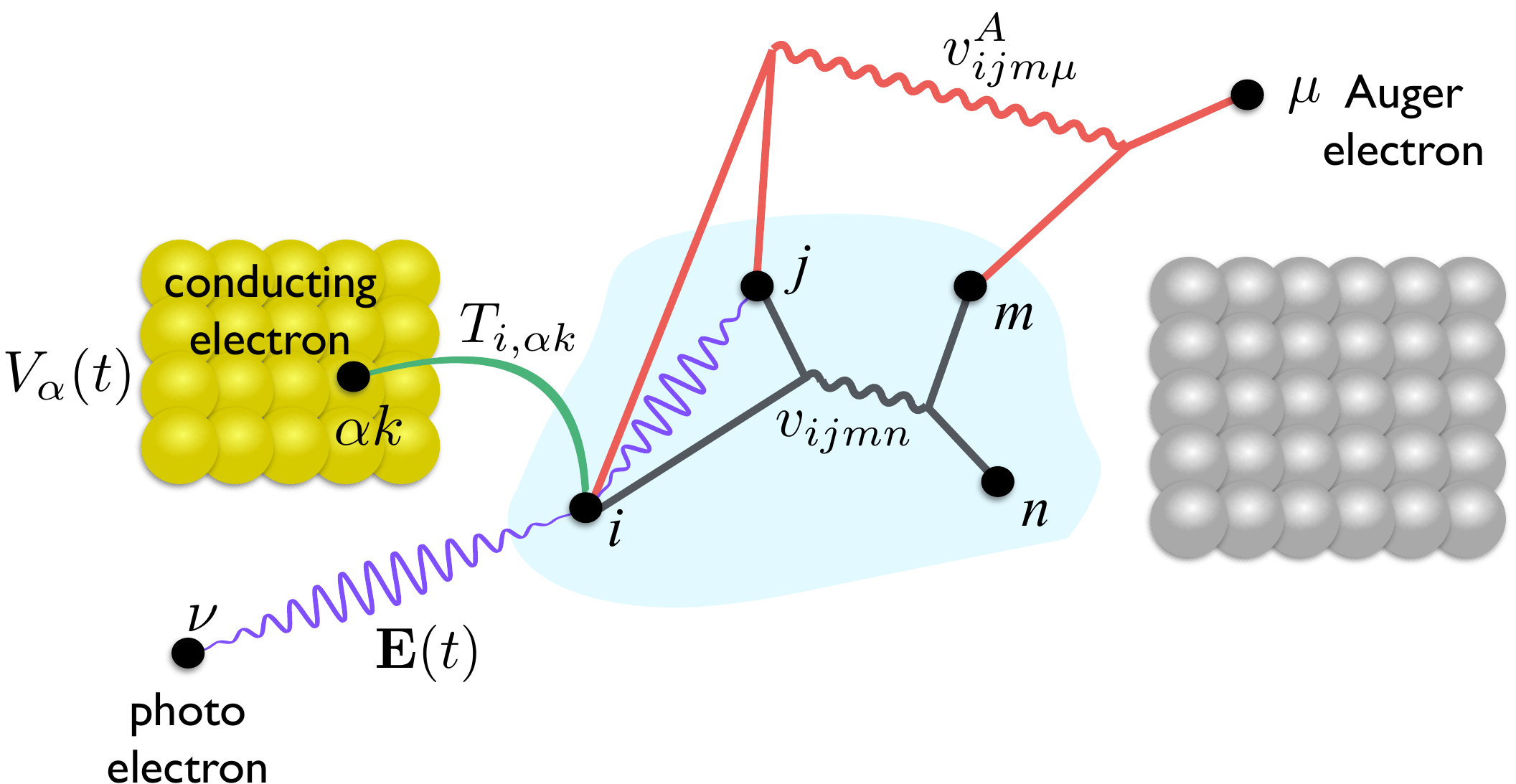}}
\caption{Illustration of the possible physical effects 
that can be addressed with the Hamiltonian in Eq.~(\ref{full-ham}).}
\label{setup}
\end{figure}

To summarize, the full Hamiltonian has the form 
\be
\hat{H}=\hat{H}_{\rm sys}+\hat{H}_{\rm cont}+
\hat{H}_{\rm lead}+\hat{H}_{\rm ion}^{\blE}+\hat{H}_{\rm 
Auger}+\hat{H}_{\rm tun}\;,
\label{full-ham}
\ee
where
\be
\hat{H}_{\rm sys}=\hat{H}_{\rm sys}^{\rm eq}+\hat{H}_{\rm sys}^{\blE}.
\ee
In Fig.~\ref{setup} we show an illustration of the physics 
that can be studied with the Hamiltonian in Eq.~(\ref{full-ham}).

\section{Green's function formulation}
\label{theorysec}

Although the full Hamiltonian in Eq.~(\ref{full-ham}) includes only the 
Auger scattering between bound electrons and continuum electrons, neglects the 
electron-electron interaction in the leads and discards the effects 
of $\blE(t)$ on the free-particle Hamiltonian of the 
continuum states, the general solution of the 
problem is still challenging. To make some progress we 
restrict the physical situations of interest. 
In experiments the momentum of the photoelectron is often well 
separated from the momentum of the Auger electron. Having in mind this 
type of experiments we take 
the set $\callS_{\rm ion}$ of photoelectron states
and the set $\callS_{\rm Auger}$
of Auger-electron states as two disjoint sets.
 This implies that 
in Eq.~(\ref{ionE-ham}) we can restrict the sum over 
$\m$  to states with $\m\in \callS_{\rm ion}$ and, similarly, in 
Eq.~(\ref{Auger-ham}) we can restrict the sum over $\m$ to 
states with $\m\in \callS_{\rm Auger}$. 

\subsection{NEGF equations}

The method of choice to investigate the electron dynamics
is the Non-Equilibrium Green's Function (NEGF) approach~\cite{danielewicz1984quantum,svl-book,balzer2012nonequilibrium}. The Green's 
function $G_{ij}(z,z')$
with times $z,z'$ on the Keldysh contour and indices $i,j$ in the 
localized sector  satisfies the 
equation of motion
\bea
\left[i\frac{d}{dz}-h_{\rm HF}(z)\right]\!G(z,z')=\d(z,z')
+\callI_{\rm emb}(z,z')
\nn\\+\callI_{\rm ion}(z,z')+
\callI_{\rm coll}(z,z')+\callI_{{\rm Auger}}(z,z')
\label{eom1}
\eea
and its adjoint. Let us discuss the quantities 
 in Eq.~(\ref{eom1}). The matrix $h_{\rm HF}$ is the 
single-particle Hartree-Fock (HF) Hamiltonian 
with elements
\be
h_{{\rm HF},ij}(z)\equiv h^{\rm eq}_{ij}+V_{{\rm 
HF},ij}(z)+\blE(z)\cdot\bld_{ij},
\label{hHF}
\ee
where the HF potential 
\be
V_{{\rm HF},ij}(z)\equiv \sum_{mn}\r_{nm}(z)w_{imnj}(z)
\label{VHF}
\ee
is expressed in terms of the one-particle density matrix 
\be
\r_{nm}(z)\equiv -iG_{nm}(z,z^{+})
\ee
and the difference between the direct and 
exchange Coulomb integrals
\be
w_{imnj}(z)\equiv  v_{imnj}(z)-v_{imjn}(z).
\label{wdef}
\ee
Hereafter we consider the more general case of a time-dependent interaction 
$v=v(z)$, useful to deal with adiabatic switching, interaction 
quenches, etc.. 

The {\em embedding}/{\em ionization} integral 
\be
\callI_{\rm emb/ion}(z,z')\equiv\int d\bar{z}\;\S_{\rm 
emb/ion}(z,\bar{z})G(\bar{z},z')
\ee
is a convolution on the Keldsyh contour between  
the embedding/ionization self-energy and the Green's function. 
The embedding self-energy is responsible for tunneling of  
electrons  to/from the leads and reads~\cite{MW92,JWM94,mssvl.2009}
\be
\S_{{\rm emb},ij}(z,\bar{z})=\sum_{\a k}T_{i,\a k}(z)
g_{\a k}(z,\bar{z})T_{\a k,j}(\bar{z}),
\label{sigma-emb}
\ee
where we allow for a time-dependent tunneling amplitude $T_{i,\a 
k}(z)$. The ionization self-energy is instead responsible for the 
photoionization of the system and reads~\cite{PUvLS.2015,PSMS.2018}
\be
\S_{{\rm ion},ij}(z,\bar{z})=\sum_{\m\in\callS_{\rm ion}}
\left(\blE(z)\cdot\bld_{i\m}\right)
g_{\m}(z,\bar{z})\left(\blE(\bar{z})\cdot\bld_{\m j}\right).
\label{sigma-ion}
\ee
Both self-energies are expressed in terms of a free-particle Green's function
$g$. For the embedding self-energy $g_{\a k}$ is the solution 
of the equation of motion
\be
\left[i\frac{d}{dz}-\F_{\a}(z)\left(\e_{\a k}-V_{\a}(z)\right)\right]g_{\a 
k}(z,\bar{z})=\d(z,\bar{z}),
\ee
whereas for the ionization self-energy $g_{\m}$ is the solution of the equation of motion
\be
\left[i\frac{d}{dz}-\e_{\m}\right]g_{\m}(z,\bar{z})=\d(z,\bar{z}).
\label{eomgmu}
\ee
All equations of motion, including Eq.~(\ref{eom1}), are solved with the 
appropriate Kubo-Martin-Schwinger boundary conditions~\cite{svl-book}. 

\begin{figure}[t]
\includegraphics[width=0.45\textwidth]{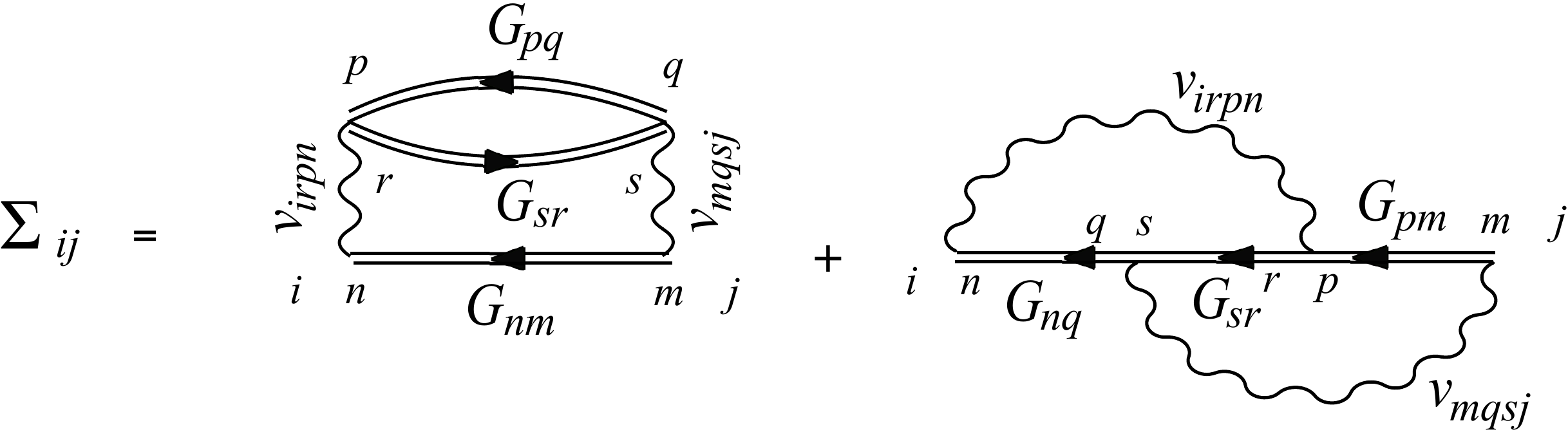}
\caption{Diagrammatic representation of the 2B self-energy. 
Wiggly lines denote the Coulomb interaction
$v$.}
\label{2Bsigma}
\end{figure}

There are two more terms to be discussed. The first term is the {\em collision} 
integral
\be
\callI_{\rm coll}(z,z')\equiv\int d\bar{z}\;\S(z,\bar{z})G(\bar{z},z'),
\label{coll-int}
\ee
where the correlation self-energy $\S=\S[v,G]$ is a functional of the 
interaction $v$ and the Green's function $G$. The exact $\S$ is the sum of all 
skeletonic self-energy diagrams with propagators $G$ and interaction 
lines $v$~\cite{svl-book}. In CHEERS this 
self-energy is implemented at the level of the second-Born (2B)
approximation, i.e.,
\bea
\S_{ij}(z,\bar{z})=
\sum_{mn\, pq\,sr}G_{mn}(z,\bar{z})G_{pq}(z,\bar{z})G_{sr}(\bar{z},z)
\nn\\
\times v_{irpm}(z)w_{nqsj}(\bar{z}).
\label{2bse}
\eea
In Fig.~\ref{2Bsigma} we show the corresponding diagrammatic 
representation. 
We mention that the 2B approximation has been successfully applied 
to equilibrium spectral properties~\cite{schuler2017} 
and total energies~\cite{DahlenLeeuwen2005} of molecular systems.
Comparisons against numerically accurate 
real-time simulations of 1D systems~\cite{PhysRevA.82.033427,BalzerHermanns2012}
and weakly correlated model nanostructures of different 
geometries~\cite{Sakkinen-2012,HermannsPRB2014,CTPPBonitz2016,HopjanPRL2016,ReichmanEPL2016,Joost2017,UKSSKvLG.2011,C60paper2018}
indicate that the 2B approximation 
remains accurate even out of equilibrium.

The last term $\callI_{{\rm Auger}}$ can be written as in Eq.~(\ref{coll-int}) but 
the self-energy contains all diagrams with at least one Auger 
interaction line $v^{A}$~\cite{PhysRevB.39.3489,PhysRevB.39.3503}. If we are only interested 
in describing the Auger physics then we can approximate
\be
\callI_{\rm Auger}(z,\bar{z})
=\int d\bar{z}\;\S_{\rm Auger}(z,\bar{z})G(\bar{z},z'),
\label{Auger-int}
\ee
where the Auger self-energy reads~\cite{Covito2018}
\bea
\S_{{\rm Auger},ij}(z,\bar{z})=
\sum_{mn\, pq}\sum_{\m\n\in\callS_{\rm Auger}}G_{mn}(z,\bar{z})
\nn\\
\times
\left[G_{\m\n}(z,\bar{z})G_{pq}(\bar{z},z)(v^{A}_{iqm\m}w^{A}_{\n 
npj}+v^{A}_{iq\m m}w^{A}_{n\n pj})
\right.
\nn\\
+\left.
G_{pq}(z,\bar{z}) G_{\m\n}(\bar{z},z)
v^{A}_{i\n pm}w^{A}_{nq\m j}\right].
\label{2bse-Auger}
\eea
This self-energy corresponds to the 2B approximation with interaction 
lines $v^{A}$ and follows from the many-body identity 
$\S=-iv^{A}G_{2}G^{-1}$ where the two-particle Green's function $G_{2}$ 
describes a single Auger scattering.
Notice that for strong valence-valence repulsive energies (typically 
well above $1$~eV) it is 
crucial to replace the (first-order) 
single-scattering approximation to  $G_{2}$ with the 
$T$-matrix approximation in the particle-particle 
channel~\cite{cini1993two,PhysRevLett.39.504}. Currently, 
CHEERS does not contain implementations of the $T$-matrix 
approximation and hence it can only be used to study Auger decays in 
molecules (e.g., 
organic molecules) with fairly delocalized valence orbitals . 

We observe that $\S_{\rm Auger}$ depends on one Green's function with 
both indices in 
the continuum. Therefore Eq.~(\ref{eom1}) for $G_{ij}$ can be solved only provided 
that we couple it to an equation for $G_{\m\n}$. To second order in 
$v^{A}$ the equation of motion for $G_{\m\n}$ with $\m,\n\in 
\callS_{\rm Auger}$ is
\bea
\left[i\frac{d}{dz}-\e_{\m}\right]G_{\m\n}(z,z')=\d(z,z')+
\sum_{\r\in\callS_{\rm Auger}}\sum_{mn\, pq\,sr}
\nn\\
\times\!\int \!d\bar{z}\;
G_{mn}(z,\bar{z})G_{pq}(z,\bar{z})G_{sr}(\bar{z},z)
v^{A}_{\m rpm}w^{A}_{nqs\r}
\nn\\
\times G_{\r\n}(\bar{z},z)
\label{eom-GAuger}
\eea
and it is linear in $G_{\m\n}$. An important consequence of this 
result is that  if the Green's 
function of the initial state is block-diagonal in the indices $i$ 
and $\m$, i.e., $G_{i\m}=0$, then it remains block diagonal.

Equations~(\ref{eom1}) and (\ref{eom-GAuger}) form a coupled system of 
nonlinear integro-differential equations and except for the 2B self-energy approximation no 
other approximations have been made. The full numerical solution of 
these equations requires to convert the contour-time $G$ to 
real-time $G$'s, a procedure leading to the so-called Kadanoff-Baym 
equations (KBE). The numerical solution of the KBE is rather demanding, 
especially for large basis sets and many nonvanishing four-index Coulomb integrals. In 
the next section we discuss how the computational cost is drastically 
reduced by the GKBA.

\subsection{GKBA equations}

We rewrite the equation of motion (\ref{eom1}) as
\bea
\left[i\frac{d}{dz}-h_{\rm HF}(z)\right]\!G(z,z')=\d(z,z')\quad\quad\quad
\nn\\+
\label{eom2}\int \!d\bar{z}\;\S_{\rm tot}(z,\bar{z})G(\bar{z},z'),
\eea
with total self-energy
\be
\S_{\rm tot}=\S_{\rm ion}+\S_{\rm emb}+\S+\S_{\rm Auger}.
\ee
In  $\S_{\rm tot}$ only the last two terms are functionals of 
$G$. Choosing $z$ and $z'$ on different branches of the Keldysh 
contour we obtain the KBE for the lesser and greater Green's 
functions~\cite{danielewicz1984quantum,svl-book,balzer2012nonequilibrium}  
\bea
\left[i\frac{d}{dt}-h_{\rm HF}(t)\right]\!G^{\lessgtr}(t,t')=
\int \!d\bar{t}\;\S^{\rm R}_{\rm tot}(t,\bar{t})G^{\lessgtr}(\bar{t},t')
\nn\\+
\label{eom2}\int \!d\bar{t}\;\S^{\lessgtr}_{\rm tot}(t,\bar{t})
G^{\rm A}(\bar{t},t'),\quad
\label{kbe1}
\eea
where, for any function $F$, the superscript ${\rm R}$ and ${\rm A}$ 
denotes the retarded and advanced 
components respectively:
\be
F^{\rm R/A}(t,t')=\pm\th(\pm t\mp t')[F^{>}(t,t')-F^{<}(t,t')].
\ee
Similarly, from the adjoint of Eq.~(\ref{eom1}) we find
\bea
G^{\lessgtr}(t,t')\left[\frac{1}{i}\frac{\overleftarrow{d}}{dt'}-h_{\rm HF}(t')\right]=
\int \!d\bar{t}\;G^{\rm R}(t,\bar{t})\S_{\rm tot}^{\lessgtr}(\bar{t},t')
\nn\\+
\label{eom2}\int \!d\bar{t}\;G^{\lessgtr}(t,\bar{t})
\S_{\rm tot}^{\rm A}(\bar{t},t').\quad
\label{kbe2}
\eea

Without any loss of generality we assume that the system is in 
equilibrium until a certain time $t_{\rm switch}>0$, hence 
$V_{\a}(t)=\blE(t)=0$ for $t<t_{\rm switch}$. To obtain the 
{\em correlated} and {\em contacted} (to leads, if any) Green's 
function we  solve 
Eqs.~(\ref{kbe1}) and (\ref{kbe2}) with: 
\begin{itemize}
    \item
 initial condition $G^{\lessgtr}(0,0)$ 
given by the HF (hence uncorrelated) lesser/greater Green's function of the uncontacted system 
\item
self-energies calculated using a TD interaction $v(t)=s(t)v$  and 
tunneling amplitude $T(t)=s(t)T$ where $s(t)$ is a 
slow and smooth switching function
between the times $t=0$ and $t=t_{\rm switch}$, typically 
$s(t)=\sin^{2}(\frac{\p t}{2 t_{\rm switch}})$. 
\end{itemize}
The time $t_{\rm 
switch}$ is therefore a convergence parameter to be chosen in such a 
way that the observables of interest are constant 
 in the absence of external fields
for times $t>t_{\rm switch}$.
This initial time-propagation serves to build up correlations 
in the inital state. 
Since $v=T=0$ for times $t<0$ 
the time integrals in the KBE run from 
$0$ to $\iif$, i.e., $\int d\bar{t}=\int_{0}^{\iif}d\bar{t}$. 

Subtracting Eq.~(\ref{kbe2}) from Eq.~(\ref{kbe1}) and setting 
$t'=t$ we obtain the equation of motion for the one-particle density 
matrix $\r_{ij}=-iG_{ij}^{<}(t,t)$
\be
\dot{\r}(t)+i\left[h_{\rm 
HF}(t),\r(t)\right]=-\callI_{\rm tot}(t)-\callI^{\dag}_{\rm 
tot}(t),
\label{eomrho}
\ee
where
\be
\callI_{\rm tot}(t)=\!
\int_{0}^{t}\! d\bar{t}\!\left[\S_{\rm 
tot}^{>}(t,\bar{t})G^{<}(\bar{t},t)-
\S_{\rm tot}^{<}(t,\bar{t})G^{>}(\bar{t},t)\right].
\label{collint-for-rho}
\ee
With similar steps, starting from Eq.~(\ref{eom-GAuger}) and its 
adjoint we can easily write down the equation of motion for the 
one-particle density matrix $f_{\m\n}(t)\equiv -iG^{<}_{\m\n}(t,t)$ with 
both indices $\m,\n\in\callS_{\rm Auger}$
\be
\dot{f}_{\m\n}(t)+i(\e_{\m}-\e_{\n})f_{\m\n}(t)=
-\callJ_{\m\n}(t)-\callJ^{\ast}_{\n\m}(t),
\label{eom-f}
\ee
to be solved with boundary conditions $f_{\m\n}(0)=0$ (no electrons 
in the continuum states at time $t=0$). The right hand 
side of Eq.~(\ref{eom-f}) describes the Auger scattering between 
localized electrons and
continuum electrons  and reads
\bea
\callJ_{\m\n}(t)=
\sum_{\r\in\callS_{\rm Auger}}\sum_{mn\, pq\,sr}\int_{0}^{t} \!d\bar{t}
\,v^{A}_{\m rpm}(t)w^{A}_{nqs\r}(\bar{t})
\nn\\
\times\!\left[
G^{>}_{mn}(t,\bar{t})G^{>}_{pq}(t,\bar{t})G^{<}_{sr}(\bar{t},t)
G^{<}_{\r\n}(\bar{t},t)\right.\;\;\;
\nn\\
\left.-
G^{<}_{mn}(t,\bar{t})G^{<}_{pq}(t,\bar{t})G^{>}_{sr}(\bar{t},t)
G^{>}_{\r\n}(\bar{t},t)\right].\;\;
\label{eomAugerrho}
\eea

Due to the implicit (through $\S$ and $\S_{\rm Auger}$)
and explicit dependence on $G^{\lessgtr}$ evaluated at times 
$t\neq\bar{t}$, Eqs.~(\ref{eomrho}) and (\ref{eom-f}) do not 
form a closed system of equations for $\r$ and $f$. 
To close the system we make the Generalized 
Kadanoff-Baym Ansatz~\cite{PhysRevB.34.6933} (GKBA) 
\begin{subequations}
\bea
-G^{<}_{ij}(t,t')\!&=&\!\sum_{m}\left[G^{\rm 
R}_{im}(t,t')\r_{mj}(t')-\r_{im}(t)G_{mj}^{\rm A}(t,t')\right],
\nn\\
\label{gkba<}
\\
G^{>}_{ij}(t,t')\!&=&\!\sum_{m}\left[G^{\rm R}_{im}(t,t')\bar{\r}_{mj}(t')
-\bar{\r}_{im}(t)G^{\rm A}_{mj}(t,t')\right],
\label{gkba>}
\nn\\
\eea
\label{gkba}
\end{subequations}

\noindent
where $\bar{\r}=1-\r$. The functional form of the retarded/advanced propagator 
$G^{\rm R/A}=G^{\rm R/A}[\r]$ will be discussed in Section~\ref{propsec}. 
For the Green's function with both indices in the continuum, 
in addition to the GKBA we  discard the off-diagonal matrix 
elements, i.e., we write
\be
G_{\m\n}=\d_{\m\n}G_{\m},
\ee
and approximate $G^{\rm R/A}_{\m}\simeq g^{\rm R/A}_{\m}$ where $g_{\m}$ 
is the solution of Eq.~(\ref{eomgmu}). Hence
\begin{subequations}
\bea
-G^{<}_{\m}(t,t')\!&=&\!
g^{\rm R}_{\m}(t,t')f_{\m}(t')-f_{\m}(t)g_{\m}^{\rm A}(t,t'),
\nn\\
\label{gkba<cont}
\\
G^{>}_{\m}(t,t')\!&=&\!g^{\rm R}_{\m}(t,t')\bar{f}_{\m}(t')
-\bar{f}_{\m}(t)g^{\rm A}_{\m}(t,t'),
\label{gkba>cont}
\nn\\
\eea
\label{gkba-Auger}
\end{subequations}

\noindent
where $\bar{f}_{\m}=1-f_{\m}$. With these approximations it is a matter of 
simple algebra to show that
\be
\callJ_{\m\n}(t)=
\int_{0}^{t} \!d\bar{t}\left[
K^{>}_{\m\n}(t,\bar{t})f_{\n}(\bar{t})
+K^{<}_{\m\n}(t,\bar{t})\bar{f}_{\n}(\bar{t})\right]
\ee
where the kernel 
\bea
K^{\lessgtr}_{\m\n}[\r](t,\bar{t})&=&i\sum_{mn\, pq\,sr}
v^{A}_{\m rpm}(t)w^{A}_{nqs\n}(\bar{t})
\nn\\
&\times&\!\!
G^{\lessgtr}_{mn}(t,\bar{t})G^{\lessgtr}_{pq}(t,\bar{t})G^{\gtrless}_{sr}(\bar{t},t)
e^{-i\e_{\m}(\bar{t}-t)}\quad
\eea
is a functional, through the GKBA, of $\r$ only. 

With Eqs.~(\ref{gkba}) and (\ref{gkba-Auger}) also the right hand side of 
Eq.~(\ref{eomrho}) becomes a functional of $\r$ and 
$f$ only. We thus obtain 
two coupled equations for the  
one-particle density matrix of an interacting system driven out of equilibrium 
by arbitrary biases $V_{\a}(t)$ and electric fields $\blE(t)$ switched on 
at times $t>t_{\rm switch}$:
\be
\left\{
\begin{array}{l}
\dot{\r}=-i\left[h_{\rm HF}[\r],\r\right]
-\callI_{\rm tot}[\r,f]-\callI^{\dag}_{\rm tot}[\r,f]
\\ \\
\dot{f}_{\m}=-\callJ_{\m\m}[\r,f]-\callJ^{\ast}_{\m\m}[\r,f]
\end{array}
\right..
\label{CHEERSeq}
\ee
These coupled equations govern the correlated electron dynamics
in the NEGF+GKBA approach and are the equations solved by the CHEERS 
code. Both $\callI(t)$ and $\callJ(t)$ depend on the density matrix at 
all previous times, thus introducing a memory dependence in the evolution.
Time-dependent HF results are  recovered by setting 
$\S=\S_{\rm Auger}=\callJ=0$. Notice that for $\S_{\rm 
Auger}=\callJ=0$ the 
equations decouple and one needs to solve only the first equation 
since $f_{\m}=0$ is a solution. 

The computational time to solve Eqs.~(\ref{CHEERSeq})
scales like $N_{t}^{2}N_{v}$ 
where $N_{t}$ is the number of 
time steps whereas $N_{v}=\max[N_{\rm 
bound}^{\mathfrak{p}},N_{\rm bound}^{\mathfrak{q}}N_{\rm cont}]$. 
Here $N_{\rm bound}$ is the dimension of the density 
matrix $\r$,  $N_{\rm cont}$ is the length of the vector $f_{\m}$ and  
the powers $3\leq \mathfrak{p}\leq 5$ and 
$2\leq\mathfrak{q}\leq 4$ depend on  how sparse
the Coulomb tensors $v_{ijmn}$ and $v^{A}_{ijmn}$ are.
In CHEERS the largest arrays are 
complex, double-precision, three-dimensional arrays of dimension 
$N_{\rm bound}\times N_{\rm bound}\times N_{t}$, and there are  
four such arrays. Thus, for a cluster with $\sim 3$ GB RAM per core CHEERS 
can perform simulations with $4 \times 16 \times N_{\rm bound}^{2}N_{t} < 3 \times 10^9$, i.e.,
$N_{\rm bound}^{2}N_{t} < 4.7 \times 10^7$.

\section{Post-processing: observable quantities}
\label{cheerspp-sec}

\begin{figure}[t]
\centerline{
\includegraphics[width=0.47\textwidth]{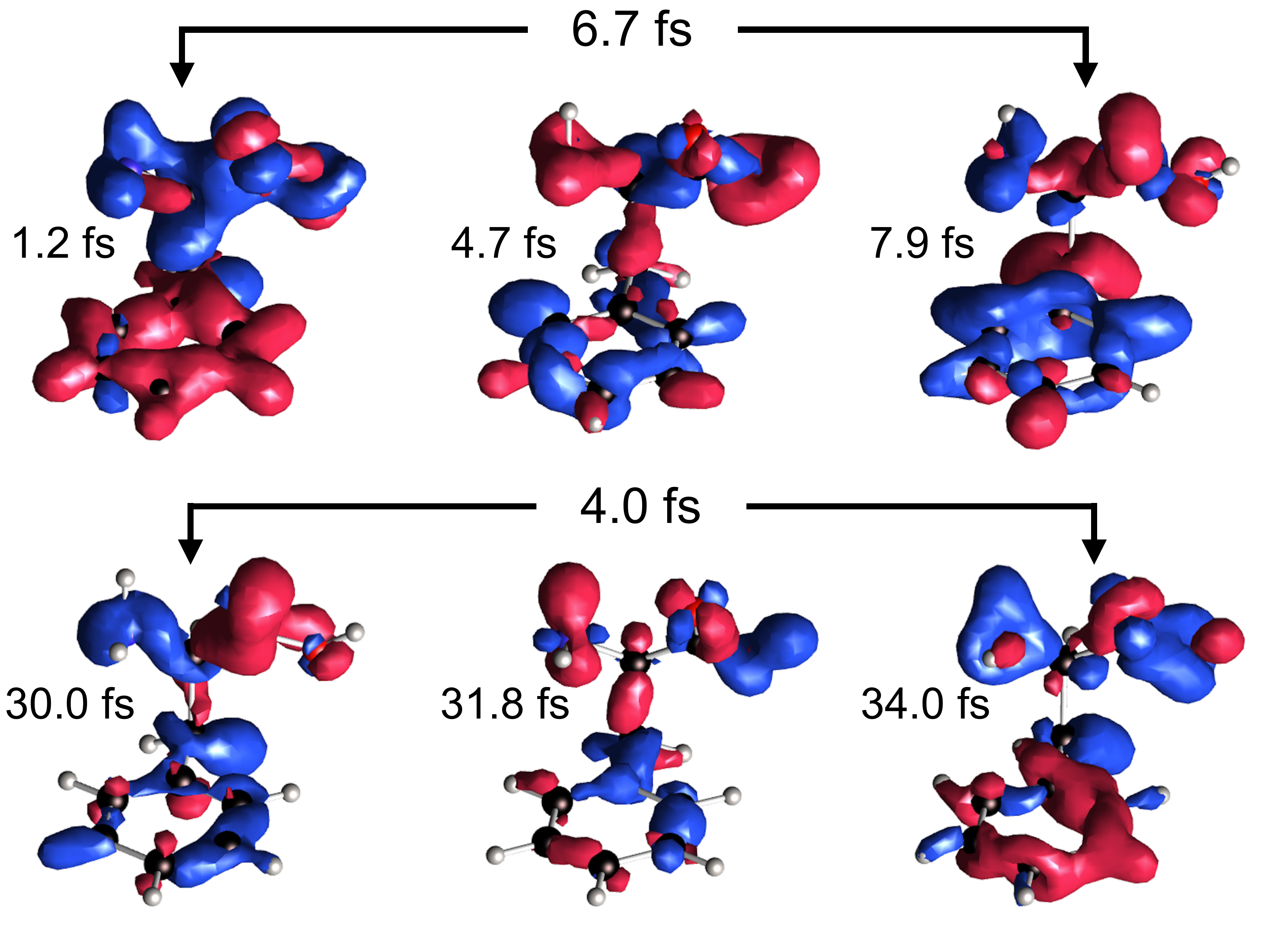}
}
\caption{Snapshots of the density variation in 
the phenylalanine aminoacid induced by an ionizing attosecond XUV 
pulse. The excess of hole density (blue) and electron density (red) 
refer to the density averaged over the full time simulation. 
Reprinted figure with permission from~\cite{PSMS.2018}. Copyright 2018 
by the American Chemical 
Society.}
\label{phenyl}
\end{figure}

From the solution of Eqs.~(\ref{CHEERSeq}) we can calculate several 
quantities of physical interest. The most straighforward ones are the 
local density
\be
n(\blr,t)=\sum_{ij}\sum_{\s}\vf_{i}^{\ast}(\blr\s)\vf_{j}(\blr\s)\r_{ji}(t),
\ee
spin density
\be
\bls(\blr,t)=\sum_{ij}\sum_{\s\s'}\vf_{i}^{\ast}(\blr\s)\bgs_{\s\s'}\vf_{j}(\blr\s')\r_{ji}(t)
\ee
local paramagnetic current
\be
\blj(\blr,t)=\frac{1}{2}\sum_{ij}\sum_{\s}
\Im
\left[\vf_{i}^{\ast}(\blr\s)\grad\vf_{j}(\blr\s)
\r_{ji}(t)\right],
\label{param-current}
\ee
and, more generally, any one-body observable. 
In Fig.~\ref{phenyl} we show the snapshots of the density variation in 
the phenylalanine aminoacid induced by an ionizing attosecond XUV 
pulse~\cite{PSMS.2018}, see Section~\ref{ks-basis-sec} for the 
implementation details.

Depending on the physical problem these basic 
quantities can be further manipulated to calculate typical  
experimental outcomes. In the following we discuss some of them.

\subsection{Quantum Transport}

In molecular electronics the Hamiltonian of the system describes a junction connecting two 
leads (a source and a drain) kept at a potential difference $V$. 
One is usually interested in the total current $I$ flowing 
through a surface $S$ perpendicular to the electron stream
\be
I(t)=\int_{S} d^{2}r \,j_{\parallel}(\blr,t),
\label{Itj}
\ee
with $j_{\parallel}$  the longitudinal component of $\blj$, see 
Eq.~(\ref{param-current}). For a 
DC bias $I(t)$ attains a steady value as $t\to \iif$ and this value 
can be used to calculate the $I-V$ characteristics or the differential 
conductance $\callG=dI/dV$. Of course, the 
full time-evolution provides other useful information. 
The characteristic time to reach a steady state and 
the frequencies of the transient oscillations are just two examples. 
Controlling these properties is crucial to engineer
 ultrafast molecular devices. 

Time-dependent potential differences $V(t)$ do 
not bring additional complications nor an increased computational 
effort. We can, for instance, superimpose an AC voltage of frequency $\W$ 
to a DC voltage, i.e., $V(t)=V_{\rm DC}+V_{\rm AC}\sin(\W t)$, and calculate 
the averaged current as well as its Fourier coefficients as functions 
of $V_{\rm DC}$, $V_{\rm AC}$ and $\W$. In addition to provide an 
alternative to Floquet schemes~\cite{Stefanuccipumping}, 
working in the time domain is particularly advantageous 
to deal with systems perturbed by multichromatic drivings. This is the case 
of, e.g.,  AC transport with superconducting leads~\cite{spc.2010} or 
optical spectra of junctions under AC voltages~\cite{GalperinCSR2017}.

The current $I(t)$ can be evaluated either at an interface passing 
through the junction, in accordance with Eq.~(\ref{Itj}), or  
at the 
interface with the $\a$
lead through the Meir-Wingreen formula~\cite{MW92,JWM94} 
\be
I_{\a}(t)=4\Re\!\int_{0}^{t}\!\!d\bar{t}\,\,
\Tr\left[
\S_{\a}^{>}(t,\bar{t})G^{<}(\bar{t},t)-\S_{\a}^{<}(t,\bar{t})G^{>}(\bar{t},t)
\right]
\label{mw-formula}
\ee
where $\S_{\a}$ is the $\a$-th contribution to the embedding 
self-energy of Eq.~(\ref{sigma-emb}) and $G^{\lessgtr}$ are 
calculated from $\r$ through the GKBA. In 
Ref.~\onlinecite{LPUvLS.2014} we showed that the GKBA results for the 
current at the interfaces are in excellent agreement with the full KBE results provided 
that the bias difference is much smaller than the bandwidth of the 
leads. We mention that an improved version of the GKBA has been recently proposed to 
deal with nontrivial spectral structures of the leads 
density-of-states~\cite{Kalvova.2017}, thus widening the potential applications of the 
GKBA in quantum transport.
 
The energy current $J_{\a}(t)$, defined as the rate of change of the energy of 
lead $\a$, can be calculated similarly to the charge current in 
Eq.~(\ref{mw-formula}). It is a matter of simple 
algebra to show that~\cite{EichPrincipiVentraVignale:14}
\be
J_{\a}(t)=4\Im\!\int_{0}^{t}\!\!d\bar{t}\,\,
\Tr\left[
\dot{\S}_{\a}^{>}(t,\bar{t})G^{<}(\bar{t},t)-\dot{\S}_{\a}^{<}(t,\bar{t})G^{>}(\bar{t},t)
\right]
\label{mw-formula}
\ee
where $\dot{\S}_{\a}^{\lessgtr}(t,\bar{t})\equiv 
\frac{d}{dt}\S_{\a}^{^{\lessgtr}}(t,\bar{t})$.

\subsection{Transient Photoabsorption Spectra}

\begin{figure}[tbp]
\includegraphics[width=8.5cm]{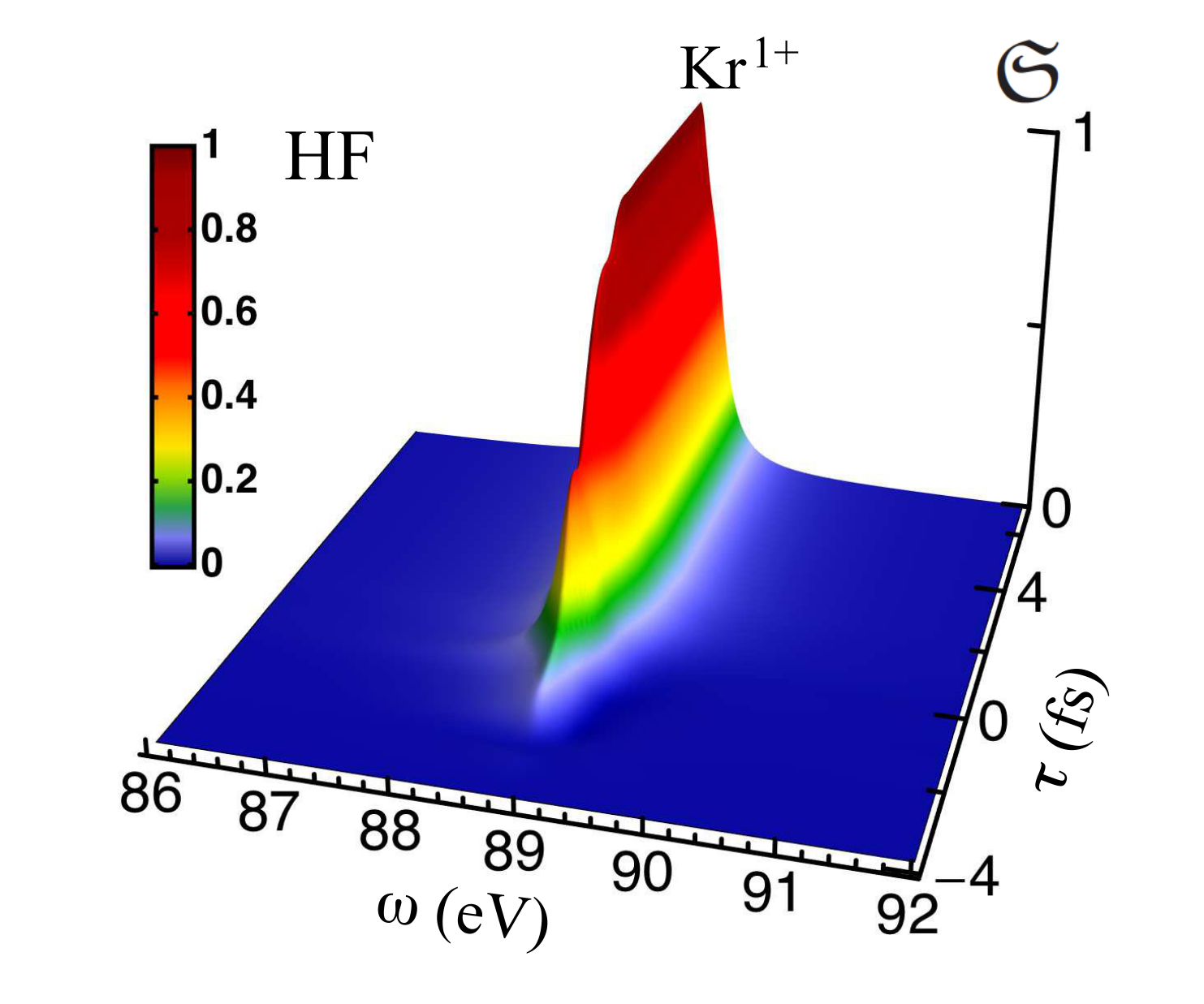}
\includegraphics[width=8.5cm]{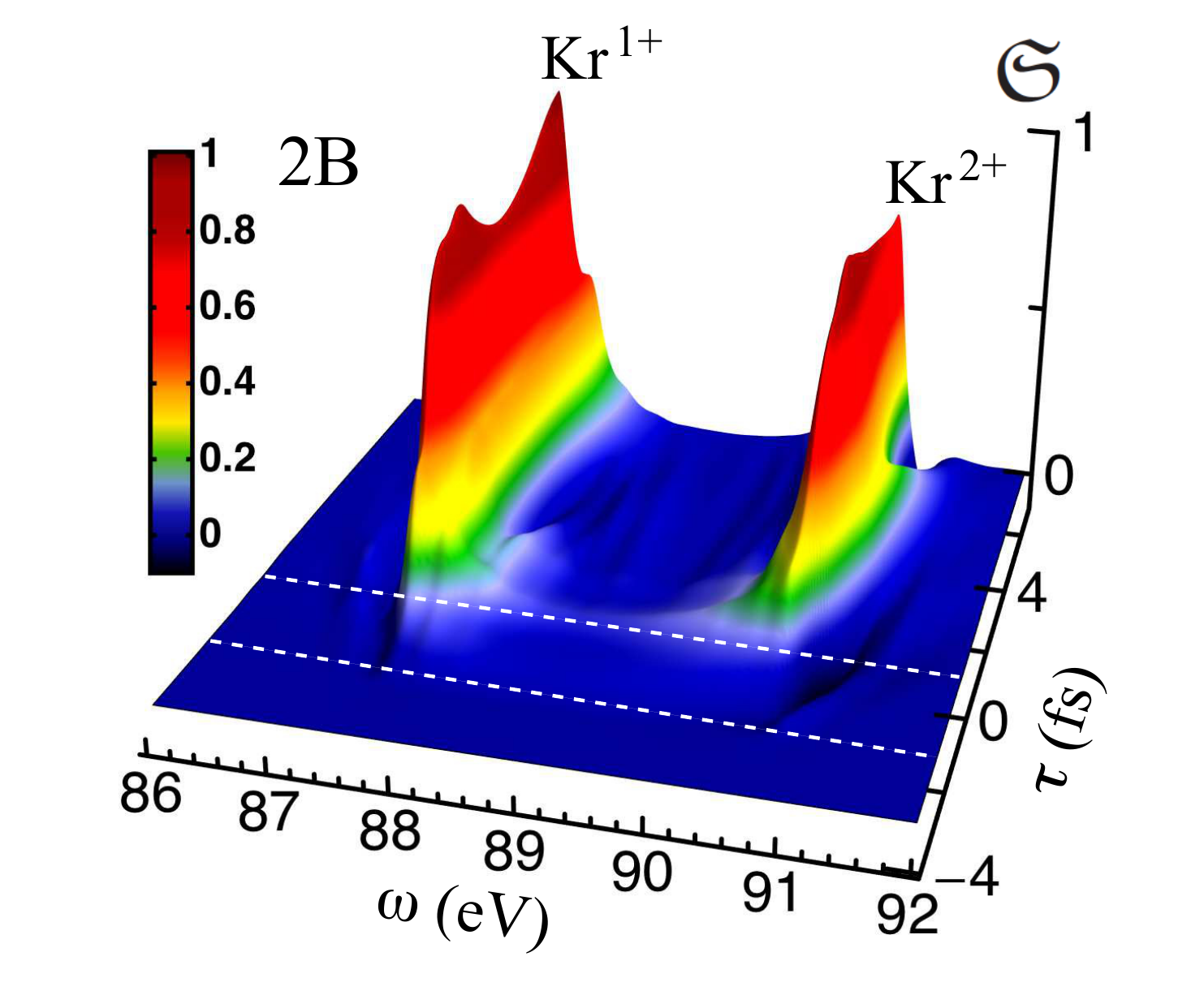}
\caption{Transient photoabsorption spectrum  (normalized to the maximum height)
of a krypton gas in the HF (top panel) and 2B 
(bottom panel) approximation. Reprinted figure with permission 
from~\cite{PUvLS.2015}. Copyright 2015 
by the American Physical 
Society.}
\label{Krtpa}
\end{figure}

In a transient photoabsorption spectrum the system is driven out of 
equilibrium by a strong laser pulse (the pump) and successively 
the intensity per unit frequency of the transmitted light of a second weak pulse (the 
probe) is measured. The resulting transient spectrum depends on the 
shape and duration of the pump and probe pulses as well as from the 
delay $\t$ between these pulses. It is therefore clear that the 
theoretical calculation of a transient spectrum calls for a 
time-dependent approach.

Let $\bra\d \bld(t,\t)\ket$ be the change of the dipole moment of the 
pump-driven system induced by an
electric probe field $\ble(t)$ impinging the system with a delay $\t$ 
with respect to the pump. 
Then the transient spectrum is given 
by
\be
\mathfrak{S}(\w,\t)=-2\Im\left[\w\,\tilde{\ble}^{\ast}(\w)\cdot\bra
\d\tilde{\bld}(\w,\t)\ket\right]
\label{nedw}
\ee
where we have used the convention that quantities with a tilde denote
the Fourier transform of the corresponding time-dependent
quantities. 

The dipole moment of the system can easily be calculated from the 
one-particle density matrix as $\bra\bld(t)\ket=\sum_{ij}\bld_{ij}\r_{ji}(t)$.
The probe-induced dipole moment $\bra\d\bld(t,\t)\ket$ is therefore 
the difference between the dipole 
moment generated by a simulation with pump and probe and the dipole 
moment generated by a simulation with only the pump. We  
observe that the pump pulse can either bring the system in an excited 
state of bound electrons  or generate a multiply ionized 
system through the ionization integral $\callI_{\rm ion}$. In the 
latter case the transient spectrum reveals features about the 
initial dynamics of the expelled photoelectrons~\cite{goulielmakis2010real}.
In Fig.~\ref{Krtpa} we display HF (top) and correlated (bottom)
calculations of the transient spectrum of a 
gas of Kr atoms initially ionized by few-cycle NIR pump and subsequently probed by an 
attosecond  XUV pulse at different delays $\t$~\cite{PUvLS.2015},
see Section~\ref{local-basis-sec} for 
the implementation details. 
In HF no sign of multiple ionization is visible. On the contrary, the 
correlated 2B results clearly show
the absorption line of Kr$^{2+}$ ions raising up a few femtosecond 
later than the absorption line of Kr$^{1+}$.

\subsection{Transient Photocurrent and Auger Current}

In the absence of leads (closed system) 
the total electric current flowing out of the system, i.e., in the continuum 
states, can be calculated from the rate of change of the total number 
of particles in the system
\be
I_{\rm ion}(t)=\frac{dN(t)}{dt}=\frac{d}{dt}\Tr[\r(t)].
\ee
If we are interested in resolving the photocurrent according to the energy of the 
photoelectrons (photoemission spectrum) we should include explicitly 
the continuum states $\m\in\callS_{\rm ion}$ in the simulation 
(instead of using $\S_{\rm ion}$).
We would then get a third equation for the occupations $f_{\m}$ with 
$\m\in\callS_{\rm ion}$ coupled to Eqs.~(\ref{CHEERSeq}).
Of course this procedure is feasible only provided that  
the energy window of the photoelectrons is not too wide and 
that the number of $\m$ states for the required energy resolution is 
not too large. At present the explicit inclusion of photoelectron states has 
been tested only in one-dimensional model systems~\cite{Covito2018}.  

The Auger self-energy accounts for processes where two
valence electrons scatter and, after the scattering,  
end up in a core state and in a 
continuum state. The rate of growth of the occupation of the continuum state 
$\m\in\callS_{\rm Auger}$ defines the Auger current 
\be
I_{\rm Auger}(t)=\frac{d}{dt}f_{\m}(t).
\ee
Through knowledge of $\r_{ij}$ and $f_{\m}$ 
we can follow in real time the Auger 
scattering and extract useful information about the Auger process, 
e.g., core relaxation time, shape of the outgoing density wavepacket, 
rearrangement of the core-excited system, etc~\cite{Covito2018}.

\section{Implementation details}
\label{imp-det-sec}

From Eqs.~(\ref{CHEERSeq}) and the definition of the various 
quantities therein we find useful to split the input parameters 
into five different groups
\begin{itemize} 
    \item {\em System}: matrix elements $h^{\rm eq}_{ij}$, 
    $\bld_{ij}$ and Coulomb integrals $v_{ijmn}$
    \item {\em Ionization}: matrix elements $\bld_{i\m}$ and energies 
    $\e_{\m}$ with $\m\in\callS_{\rm ion}$
    \item {\em Auger}: matrix elements $\bld_{i\m}$, energies 
    $\e_{\m}$ and  Coulomb integrals $v^{A}_{ijm\m}$  with 
    $\m\in\callS_{\rm Auger}$
    \item {\em Leads}: tunneling amplitudes $T_{i,\a k}$ and energies 
    $\e_{\a k}$
    \item {\em External fields}: electric field $\blE(t)$, bias 
    $V_{\a}(t)$ and temperature $T_{\a}(t)$.
\end{itemize}

\begin{figure*}[t]
\centerline{
\includegraphics[width=0.9\textwidth]{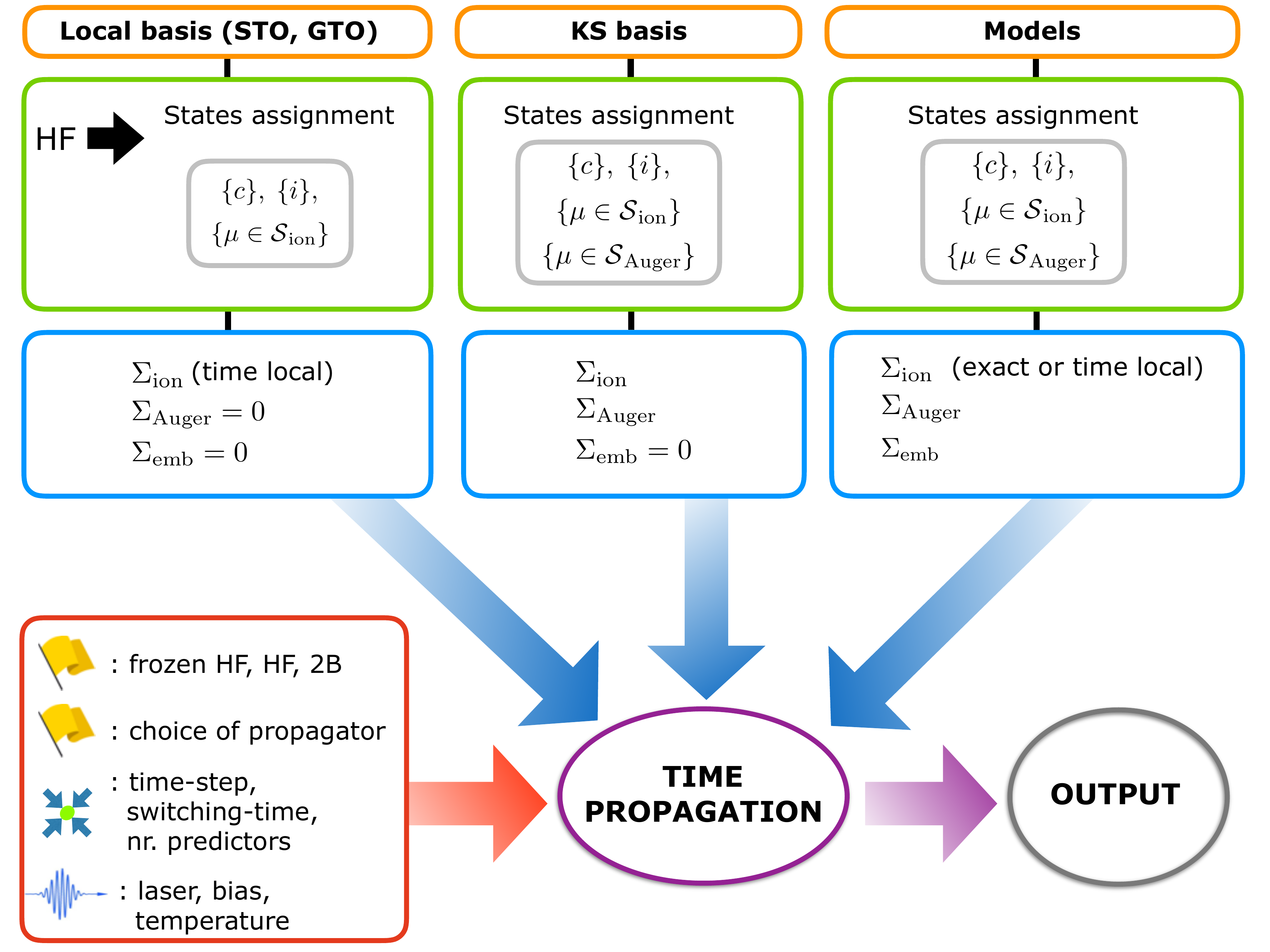}}
\caption{Architecture of CHEERS. The input contained in {\em System}, {\em Ionization}, 
{\em Auger} and {\em Leads} can be either generated from first-principles 
calculations in a localized basis (left) or KS basis (middle) or, 
alternatively, it can be  set manually for model system calculations 
(right). The level of correlation, type of propagator, convergence 
parameters and the input of {\em External fields} is given in the red 
 box (bottom right). During the time propagation the density matrix is either saved 
or processed to generate the output of interest.}
\label{flowchart}
\end{figure*}

Currently, real-time simulations in the presence of leads (open 
systems) are 
performed only for model Hamiltonians. Here, the 
input parameters are set manually and can be varied at will. 
Real-time simulations 
based on first-principles input parameters
are possible for closed systems ($\S_{\rm emb}=0$)
like atoms and molecules in external laser fields. CHEERS handles
the input parameters in 
different way depending on the nature of the 
single-particle basis set. 
In the following two subsections we describe how CHEERS 
processes the input generated in a basis of localized orbitals like, 
e.g., Slater Type Orbitals (STO) or Gaussian Type Orbitals (GTO), 
and in a basis of Kohn-Sham orbitals. 
In all cases the first step of CHEERS is to obtain 
the information contained in 
{\em System}, {\em Ionization}, 
{\em Auger} and {\em Leads}. With this information CHEERS
calculates all self-energies and then passes them to the 
time-propagation routine, see blue arrows in Fig.~\ref{flowchart}.

As illustrated in the left red box of Fig.~\ref{flowchart} the 
time-propagation routine needs also other information.
Two main flags specify the type of evolution.
One flag establishes the level of correlation: it can be either 
a HF evolution -- with the extra option of 
using the HF Hamiltonian $h_{\rm HF}[\r(t_{\rm switch})](t)$ with 
frozen $\r$ 
instead of  $h_{\rm HF}[\r(t)](t)$ -- or a correlated 2B evolution. Another flag 
sets which propagator $G^{\rm R}[\r]$ is used in the GKBA, see 
subsection~\ref{propsec} for the possible choices. Prior to the  
time evolution we also specify a few convergence parameters. The most 
important ones are the time-step, the switching time $t_{\rm switch}$ 
for the correlation build-up and the number of predictor correctors 
for each time step. Finally, we specify the driving fields in the input {\em External fields}. 
There are no restrictions on the time-dependent functions $\blE(t)$, 
$V_{\a}(t)$ 
and $T_{\a}(t)$, and the computational effort does not depend on the 
choice of these 
functions.  During the time stepping the density matrix $\r_{ij}$ and $f_{\m}$ 
are either saved or processed to generate the output described in 
Section~\ref{cheerspp-sec}.

\subsection{Localized basis}
\label{local-basis-sec}

For a description in terms of $N$ single-particle localized states $\{|i\ket\}$ 
like, e.g., the STO or GTO states,  CHEERS needs 
the matrix elements of the 
equilibrium Hamiltonian $h^{\rm eq}_{ij}$, dipole vector $\bld_{ij}$, 
overlap matrix $S_{ij}=\bra i|j\ket$ and Coulomb integrals 
$v_{ijmn}$. The first step of CHEERS  is to orthonormalize the basis according 
to $|i\ket\to \sum_{m}|m\ket S^{-1/2}_{mi}$, calculate 
$h^{\rm eq}_{ij}$, $\bld_{ij}$ and $v_{ijmn}$ in the new orthonormal basis and run a HF 
self-consistent calculation. 
The second step is 
to calculate $h^{\rm eq}_{ij}$, $\bld_{ij}$ and $v_{ijmn}$ in the
HF basis  which, by definition, diagonalizes the HF Hamiltonian $h_{{\rm HF},ij}=\e^{\rm 
HF}_{i}\d_{ij}$. This is done only for HF states with 
energy $\L_{\rm min}<\e^{\rm HF}_{i}<\L_{\rm max}$ where $\L_{\rm 
min}$ and $\L_{\rm max}$ are two convergence cutoff parameters. Of 
course if $\L_{\rm min}$ is smaller than the minimum HF eigenvalue and 
$\L_{\rm max}$ is larger than the maximum HF eigenvalue then all $N$ HF states 
are included. On the contrary, states $i=c$ of energy 
$\e_{c}^{\rm HF}<\L_{\rm min}$ are treated as core states, i.e., 
$\r_{cc'}=\d_{cc'}$. Then CHEERS calculates only $v_{iccj}$ and 
$v_{icjc}$  and adds to the equilibrium Hamiltonian the 
HF potential generated by the frozen core (fc) electrons: 
\be
h^{\rm eq}_{ij}\to
h^{\rm eq+fc}_{ij}=
h^{\rm eq}_{ij}+\sum_{c}(v_{iccj}-v_{icjc}).
\ee
The HF states $i=\m$ with energy $\e^{\rm HF}_{\m}>\L_{\rm max}$ 
are treated as noninteracting and considered as states of 
the continuum. Accordingly, CHEERS calculates only the dipole matrix elements $\bld_{i\m}$ 
later used to 
construct the ionization self-energy 
of Eq.~(\ref{sigma-ion}). 
The separation of the HF states  
is illustrated in the left green box of
Fig.~\ref{flowchart}.

After this preliminary treatment the size of the one-particle density matrix 
$\r$ in Eq.~(\ref{eomrho}) becomes $N_{\rm bound}=N-N_{c}-N_{\rm ion}$, where
$N_{c}$ is the number of HF core states and 
$N_{\rm ion}$ is the number of $\m$-states.
The STO or GTO description of the continuum is, in general, too poor for 
simulating  Auger scattering processes which are therefore not 
included, see left blue box of
Fig.~\ref{flowchart}. This 
means that for a localized basis  
CHEERS solves only the first of Eqs.~(\ref{CHEERSeq}). For problems 
involving Auger scattering see next Section.

The $\m$-states of the localized basis are used to construct an 
approximate ionization self-energy, thus accounting for possible 
photoelectrons due to external laser fields. Of course, if 
the expected energy range of the photoelectrons is not
covered by the $\e_{\m}^{\rm HF}$ then the ionization rate is  
severely underestimated. Meaningful real-time simulations do therefore 
require that the laser frequency is at least smaller than 
$\max{\{\e_{\m}^{\rm HF}\}}-2\p/T$ where $T$ is the duration of the 
ionizing pulse. 

Let us discuss the ionization self-energy. In general 
the STO or GTO basis returns only a few 
continuum states; consequently, a photoelectron would be soon reflected back.
This difficulty can be overcome provided that the ionizing laser 
pulse is well centered around some frequency $\w_{P}$. From 
Eq.~(\ref{sigma-ion}) the lesser part of  $\S_{\rm ion}$ vanishes 
whereas the greater part is given by
\be
\S_{\rm ion}^{>}(t,t')=
\sum_{ab}E_{a}(t)\s^{ab}(t-t')E_{b}(t'),
\label{selfion>}
\ee
where $E_{a}$ is the $a$-th component of the electric 
field $\blE=(E_{x},E_{y},E_{z})$ and the tensor 
\be
\s^{ab}_{ij}(t-t')\equiv -i
\sum_{\m\in\callS_{\rm ion}}
d^{a}_{i\m}\,e^{-i\e_{\m}^{\rm HF}(t-t')}\,d^{b}_{\m j},
\label{cselfion>}
\ee
depends exclusively on the matrix elements of the components $d^{a}$ 
of the dipole moment $\bld=(d^{x},d^{y},d^{z})$. The Fourier 
transform of $\s^{ab}_{ij}$ reads
\bea
\tilde{\s}^{ab}_{ij}(\w)\!\!&=&\!\!-2\p i\sum_{\m\in\callS_{\rm ion}}
d^{a}_{i\m}\,\d(\w-\e_{\m}^{\rm HF})\,d^{b}_{\m j}
\nn\\
\!\!&\approx&\!\!
2i\sum_{\m\in\callS_{\rm ion}}
d^{a}_{i\m} \Im\left[\frac{1}{\w-\e_{\m}^{\rm HF}+i\eta}\right]d^{b}_{\m j},
\label{sigmaion}
\eea
where $\eta$ is a positive constant of the order of the level spacing 
of the $\m$-states. Since $\blE(t)$ oscillates at the frequency 
$\w_{P}$ the ionization self-energy  
 is dominated by those terms in $\s(t-t')$ that 
oscillate at energy $\e_{\m}^{\rm HF}\simeq \w_{P}$. 
We do therefore implement a frequency-independent approximation
$\tilde{\s}^{ab}_{ij}(\w)\approx 
\tilde{\s}^{ab}_{ij}(\w_{P})$,
which in real time implies 
$\s^{ab}_{ij}(t-t')=\tilde{\s}^{ab}_{ij}(\w_{P})\d(t-t')$. 
Substituting this result into Eq.~(\ref{selfion>}) we get 
\be
\S_{\rm 
ion}^{>}(t,t')=-i\d(t-t')\G_{\rm ion}(t),
\label{selfion>local}
\ee
where
\be
\G_{{\rm ion},ij}(t)=i\sum_{ab}E_{a}(t)\tilde{\s}^{ab}_{ij}(\w_{P})E_{b}(t)
\label{gammaion}
\ee
is a self-adjoint positive-definite matrix for all times $t$. Thus, 
the approximate $\S_{\rm ion}$ is a local function of time as indicated 
in the left blue box of Fig.~\ref{flowchart}. 

The transient 
photoabsorption spectrum of the Kr gas
in Fig.~\ref{Krtpa}~\cite{PUvLS.2015} has been 
calculated using 
the 66 STO from Ref.~\cite{BUNGE1993113} as basis, 
generating the input in {\em System} and {\em Ionization} 
with the SMILES package~\cite{smiles1,smiles2}, freezing all electrons below the 
3$d$ shell and constructing $\S_{\rm ion}$ with the HF states of 
positive energy.

\subsection{Kohn-Sham basis}
\label{ks-basis-sec}

In general the finite system of interest can be described in terms of 
a single-particle basis formed by 
core states and a rest. The rest is a set of {\em active} states, 
i.e., states with a  
population different from 0 or 1 because of dynamical correlations or 
thermal fluctuations or external fields. Let $\callC$ and 
$\callA$ be the set of core states and active states respectively. 
Since, by definition, for $i\in\callC$ every physically relevant many-body state is an 
eigenstate of $\hat{c}^{\dag}_{i}\hat{c}_{i}$ with 
eigenvalue 1, we can work in the truncated Hilbert space of 
many-body states having the core states entirely filled. In this truncated 
Hilbert space the density matrix $\r$ satisfies again 
Eq.~(\ref{CHEERSeq}) but with a different 
HF Hamiltonian $h_{\rm HF}$. 

To determine the HF Hamiltonian for the ``active'' electrons
let us  split the 
contributions (core and active) to the Hartree and exchange potential
\bea
V_{{\rm H},ij}^{\callS}[\r]&\equiv& \sum_{mn\in\callS}v_{imnj}\r_{nm},
\quad\quad ij\in\callA
\\
V_{{\rm x},ij}^{\callS}[\r]&\equiv& -\sum_{mn\in\callS}v_{imjn}\r_{nm}
,
\quad\quad ij\in\callA.
\label{vx}
\eea
where $\callS=\callC,\callA$ and the indices $i,j$ run in the active set 
$\callA$. Taking into account that $\r_{nm}=\d_{nm}$ 
for $n,m\in\callC$ the equilibrium HF 
Hamiltonian in Eq.~(\ref{hHF}) can be rewritten as
\be
h_{{\rm HF}}[\r]=h^{\rm eq+fc}
+V_{{\rm H}}^{\callA}[\r]+V_{{\rm x}}^{\callA}[\r],
\label{hqp2}
\ee
where 
\be
h^{\rm eq+fc}=h^{\rm eq}+V_{{\rm H}}^{\callC}+V_{{\rm x}}^{\callC}
\ee
is the one-particle Hamiltonian plus the HF potential generated by 
the frozen core electrons.

So far we have not yet specified the single-particle basis.
We here consider the case of a Kohn-Sham (KS) basis. Hence we 
assume that electrons in the KS core orbitals remain frozen and 
do not participate to the dynamics. 
The equilibrium KS one-particle density matrix in the KS basis reads $\r_{{\rm 
KS},nm}=\d_{nm}$ and the corresponding equilibrium KS Hamiltonian 
is diagonal and reads
\be
h_{{\rm KS}}=h^{\rm eq}+V_{{\rm H}}^{\callC}+
V_{{\rm xc}}+V_{{\rm H}}^{\callA}[\r_{{\rm 
KS}}],
\ee
where $V_{\rm xc}$ is the exchange-correlation potential of Density 
Functional Theory (DFT). In general, $V_{{\rm H}}^{\callC}+
V_{{\rm xc}}$  is given by the sum 
of the pseudopotential and the
xc potential generated by the active electrons.
A comparison with Eq.~(\ref{hqp2}) allows us to express 
$h^{\rm eq+fc}$ in terms of the KS Hamiltonian according to
\be
h^{\rm eq+fc}=h_{{\rm KS}}-V_{{\rm xc}}-
V_{{\rm H}}^{\callA}[\r_{{\rm KS}}]+
V_{{\rm x}}^{\callC}.
\label{hqp3}
\ee

Depending on the system and laser 
pulse properties the electrons in 
states with 
$\e^{\rm KS}_{i}<\L_{\rm max}$ are explicitely propagated through 
$\r$, whereas states $i=\m$ with energy  $\e^{\rm KS}_{\m}>\L_{\rm max}$
are either assigned to $\callS_{\rm ion}$ or $\callS_{\rm Auger}$, 
see middle green box in Fig.~\ref{flowchart} 
(here $\L_{\rm max}$ is a convergence parameter).
Thus, CHEERS needs 
the KS eigenvalues $\e^{\rm KS}_{i}$ (needed to construct the KS Hamiltonian $h_{{\rm 
KS},ij}=\d_{ij}\e^{\rm KS}_{i}$),
the matrix elements
$V_{{\rm xc},ij}$, $\bld_{ij}$ and the Coulomb integrals $v_{ijmn}$
(needed to evaluate the Hartree potential $V_{{\rm 
H}}^{\callA}[\r_{{\rm KS}}]$ generated by the active KS electrons  
as well as the functionals
$V_{{\rm H}}^{\callA}$, $V_{{\rm x}}^{\callA}$ and  $\S$), 
the dipole matrix elements $\bld_{i\m}$ and KS energies 
$\e^{\rm KS}_{\m}$ 
with $\m\in \callS_{\rm ion}$
(needed to calculate the ionization self-energy $\S_{\rm ion}$) and
the Coulomb integrals $v^{A}_{ijm\m}$ and KS energies $\e^{\rm KS}_{\m}$ 
with $\m\in \callS_{\rm Auger}$ (needed to calculate the Auger self-energy 
$\S_{\rm Auger}$ and the kernel $K$), see 
middle blue box in Fig.~\ref{flowchart}. 
This input contains the necessary quantities to construct the 
functionals $\callI_{\rm tot}$ and $\callJ$, see 
Eqs.~(\ref{CHEERSeq}),  as well as  
the HF Hamiltonian in Eq.~(\ref{hqp3}). 
In fact, the only remaining unknown is $V_{{\rm x}}^{\callC}$ which, 
however, is usually small and can be neglected. 
Of course, a non-negligible $V_{{\rm x}}^{\callC}$ does not introduce  extra 
complications for the CHEERS simulations. One could  estimate 
this quantity by performing an all-electron KS calculation 
without pseudopotentials. 

The snapshots of the density variation of 
the phenylalanine aminoacid in Fig.~\ref{phenyl}~\cite{PSMS.2018} 
has been calculated by performing a DFT calculation with the Quantum 
Espresso package~\cite{QuantumEspresso} using 
norm-conserving Troullier-Martins pseudopotentials~\cite{PhysRevB.43.1993} and the PBE 
approximation~\cite{PhysRevLett.77.3865} for $V_{\rm xc}$. The 
resulting KS states have then been used to extract the matrix 
elements $V_{{\rm xc},ij}$, $\bld_{ij}$, $\bld_{i\m}$ and the Coulomb 
integrals $v_{ijmn}$ using the Yambo code~\cite{MARINI20091392}
(no Auger scattering was included since only 
valence electrons are ionized by the  XUV pulse).

\subsection{An application to Argon: STO versus KS basis}

As long as the single-particle basis is complete the CHEERS results are 
independent of the basis. 
The purpose of this Section is to illustrate this fact with an 
example. We consider the Argon 
atom and monitor the time-evolution of the occupations of the 
HF orbitals after a sudden ionization. The calculations 
are performed in two different basis:
\\
(i) the STO basis of Clementi-Roetti~\cite{ClementiRoetti} 
consisting of 32 basis functions. The input has been 
obtained with the SMILES package~\cite{smiles1,smiles2}
and no electron has been frozen (hence electrons of the K and L shells 
participate to the dynamics).
\\
(ii) the KS basis obtained by performing a DFT calculation with the 
Octopus code~\cite{OCTOPUS} using norm-conserving Troullier-Martins 
pseudopotentials~\cite{PhysRevB.43.1993} and the Perdew-Zunger xc 
functional~\cite{PZ81}.

In both cases we start from an initial density matrix 
corresponding to the state of the system just after an ionizing laser 
pulse has passed through the atom. Typical attosecond pulses remove 
less than 1$\%$ of charge from the neutral system. Here, in order to 
highlight the effects of correlations, we consider
an initial density matrix 
$\r_{a\s,b\s'}(0)=\d_{\s\s'}\r_{a,b}(0)$ where
$\r(0)=\r^{\rm eq}-\d\r$ and $\d\r$ in HF basis 
reads $\d\r_{3s,3s}=0.1$, $\d\r_{3p_{x},3p_{x}}=0.1$ and
$\d\r_{3s,3p_{x}}=\d\r_{3p_{x},3s}=-0.1$. This corresponds to remove 
0.2 electrons of spin up and down.

\begin{figure}[t]
\centerline{
\includegraphics[width=0.49\textwidth]{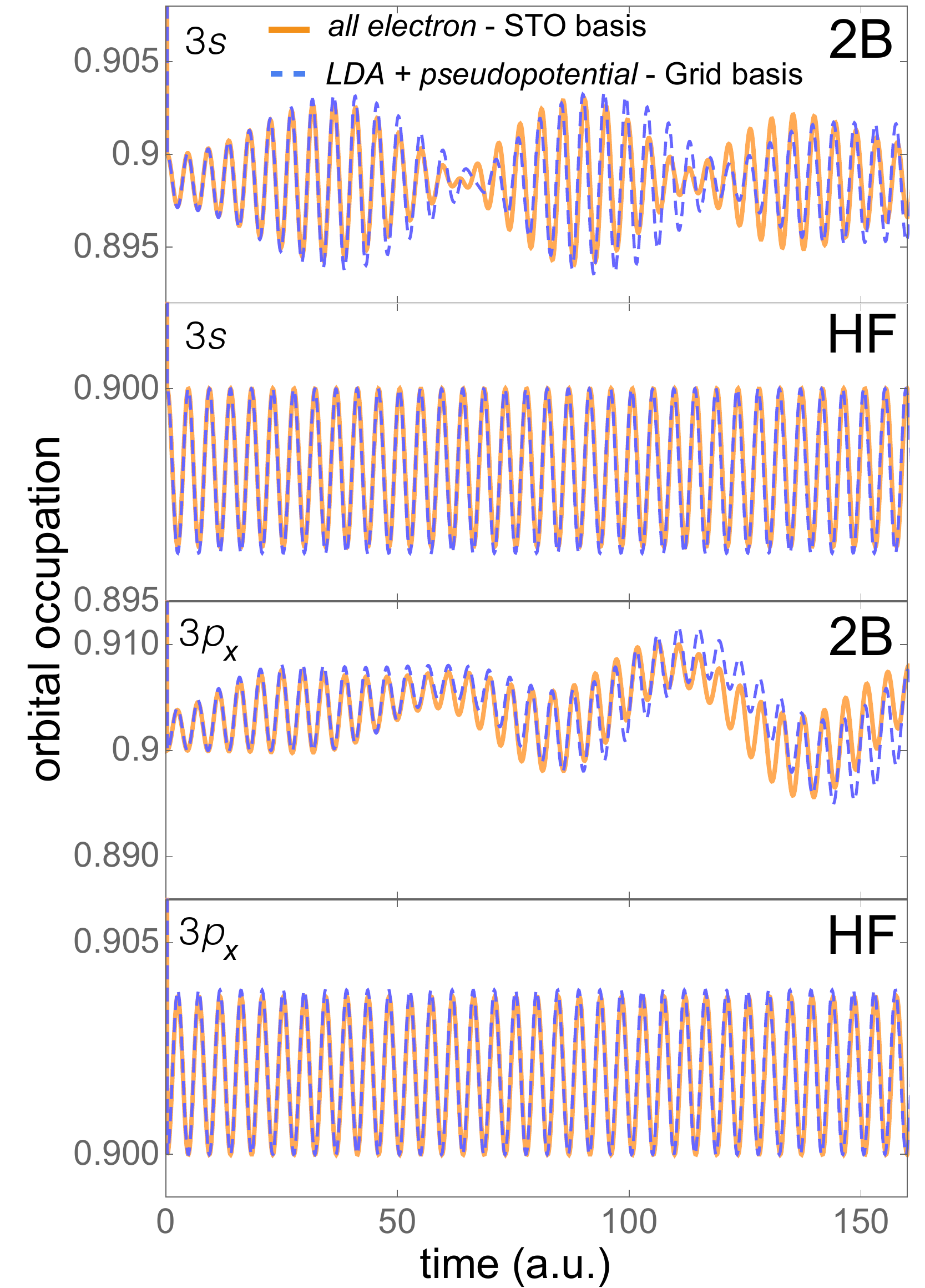}}
\caption{Time-dependent evolution of the occupations of 
the equilibrium  HF orbitals $3s$ and $3p_{x}$ after a sudden 
ionization, 
as described in the main text. The calculations have been performed 
using the HF (second and fourth panels) and 2B (first and third panels)
approximation.}
\label{argon}
\end{figure}

In Fig.~\ref{argon} we 
compare the results in the two different basis as obtained by running 
CHEERS in the HF and 2B approximation. In HF the time-evolution is 
dominated by $3s\leftrightarrow 3p_{x}$ transitions and resembles the 
evolution of a noninteracting two-level system, in agreement with the 
fact that the HF theory is a single-particle theory and the system is weakly 
correlated. On the contrary, the correlated 2B evolution highlights 
the occurrence of scatterings involving $3p_{y}$ and $3p_{z}$ 
electrons. In fact, the initial density matrix describes a 
mixture of charge neutral Ar and multiply ionized  Ar$^{n+}$ with 
$n=1,\ldots,6$. In the 
considered Hilbert space Ar$^{+}$ can only give rise to the oscillation 
corresponding to the transition $3s\leftrightarrow 3p$ while 
Ar$^{2+}$ can only give 
rise to oscillations corresponding to the transitions 
$3s^{2}\leftrightarrow 3s3p$ and $3s3p\leftrightarrow 3p^{2}$. 
These are degenerate in HF although in reality they should not. 2B correctly 
removes the degeneracy giving rise to the observed beating. 
Doubly and multiply ionized Ar atoms contributes less since we have 
removed only 40$\%$ of an electron.
Aside from the physical interpretation of the results, the figure 
clearly show 
that the outcomes stemming from using two different basis (and 
procedures) are in a fairly good agreement.

\subsection{Retarded propagator}
\label{propsec}

In this section we discuss the possible choices of the retarded 
Green's function. The exact equation of motion for $G^{\rm R}$ reads
\be
\left[
i\frac{d}{dt}-h_{\rm HF}(t)\right]\!G^{\rm R}(t,t')=
\d(t,t')+\int \!\! d\bar{t}\,\S^{\rm R}_{\rm tot}(t,\bar{t})G^{\rm 
R}(\bar{t},t')
\label{eqgrint}
\ee
to be solved with boundary condition $G^{\rm R}(t,t^{+})=-i$. 
The lowest order (in the Coulomb integrals) approximation for $G^{\rm R}$ is obtained by setting  $\S^{\rm R}_{\rm 
tot}=0$. In this case we get the HF propagator
\be
G^{\rm R}(t,t')=-i\th(t-t')\,T\,e^{-i\int_{t'}^{t}d\bar{t}\,h_{\rm 
HF}(\bar{t})}.
\label{HFGR}
\ee
In CHEERS all approximations to $G^{\rm R}$ have the form of 
Eq.~(\ref{HFGR}) where $h_{\rm HF}$ is replaced by some 
quasi-particle Hamiltonian $h_{\rm qp}=h_{\rm HF}+\D$. 
For $\D=0$ we recover 
the HF propagator. The advantage of approximations like in 
Eq.~(\ref{HFGR}) is that for small $\d t$
\be
G^{\rm R}(t+\d t,t')\simeq e^{-i\frac{h_{\rm qp}(t+\d 
t)+h_{\rm qp}(t)}{2}\d t}G^{\rm R}(t,t')
\ee
and hence the calculation of $G^{\rm R}(t,t')$ for all $t'<t$ scales 
linearly in $t$. Consequently, the overall scaling remains quadratic with the 
maximum propagation time.

The presence of a continuum due to leads and/or photoelectron states 
can be partially taken into account by approximating 
\be
\S^{\rm R}_{\rm emb/ion}(t,t')\simeq -(i/2)\d(t-t')\G_{\rm emb/ion}(t)
\label{wblionemb}
\ee
where $\G_{\rm ion}$ is defined in Eq.~(\ref{gammaion}) whereas
\bea
\G_{{\rm emb},ij}=-2\sum_{k\a}
T_{i,\a k}\,
\Im\left[\frac{1}{-\e_{\a k}+i\eta}\right]T_{\a k,j},
\label{gammaemb}
\eea
see Ref.~\cite{LPUvLS.2014}. 
Setting $\S^{\rm tot}= 
\S^{\rm R}_{\rm emb}+\S^{\rm R}_{\rm ion}$ in Eq.~(\ref{eqgrint}) one 
finds $\D=-(i/2)(\G_{\rm emb}+\G_{\rm ion})$.

Correlation effects in the propagator can be taken into account
by making the approximation~\cite{LPUvLS.2014}
\bea
\int \!\! d\bar{t}\,\S^{\rm R}(t,\bar{t})G^{\rm 
R}(\bar{t},t')&\simeq& \left[\int \!\! d\bar{t}\,
\S^{\rm R}(t,\bar{t})\right]G^{\rm R}(t,t')
\nn\\
&\equiv& \tilde{\S}(t)G^{\rm R}(t,t'),
\label{rhsint}
\eea
which amounts to add $\tilde{\S}$ to $h_{\rm HF}$.
We evaluate $\tilde{\S}(t)$ in Eq.~(\ref{rhsint}) using the GKBA and 
the adiabatic propagator
\be
\tilde{G}^{\rm 
R}_{\rm ad}(t,t')=\int\frac{d\w}{2\p}\frac{e^{-i\w(t-t')}}{\w-h_{\rm 
qp}(t)+i\eta}.
\label{tildegret}
\ee
In this way we generate a self-consistent equation for 
$\tilde{\S}(t)=\tilde{\S}[\r(t),h_{\rm qp}(t)]$. In practice 
at the $n$-th time step we 
determine $\r(t_{n+1})$ from Eq. (\ref{CHEERSeq}), then calculate 
$h_{\rm qp}(t_{n+1})$ using $\tilde{\S}(t_{n+1})=\tilde{\S}(t_{n})$, 
hence $\tilde{G}^{\rm R}(t_{n+1},t')$ and finally the new $\tilde{\S}(t_{n+1})$. 
This procedure is repeated a few times to achieve 
convergence.
We point out that propagator used in the evaluation of $\callI_{\rm 
tot}$ is $G^{\rm R}$ and not the adiabatic $\tilde{G}^{\rm R}$. 
The latter is only an auxiliary quantity to evaluate 
$\tilde{\S}(t)$. 

In CHEERS the quasi-particle Hamiltonian used for $G^{\rm R}$ reads
\be
h_{\rm qp}=h_{\rm HF}-(i/2)(\a_{\rm emb}\G_{\rm emb}+\a_{\rm ion}\G_{\rm ion})
+\a_{\rm ad}\tilde{\S},
\ee
where the parameters $\a_{\rm emb},\a_{\rm ion},\a_{\rm ad}$ can be 
set to 
either $0$ or $1$.

\section{Concluding remarks}
\label{conclusionssec}

To the best of our knowledge, CHEERS is currently the only code which 
combines ab initio methods with NEGF to calculate the nonequilibrium 
dynamics of molecular systems.   
CHEERS has already been used to study 
the charge dynamics of molecular junctions~\cite{LPUvLS.2014}, time-resolved 
photoabsorption of noble gas atoms~\cite{PUvLS.2015,PUvLS.2016}, charge separation in 
donor-acceptor complexes~\cite{C60paper2018}, charge migration in 
organic molecules~\cite{PSMS.2018} and time-resolved Auger 
decays~\cite{Covito2018}. The code handles inputs obtained in
any basis and can perform all-electrons as well as pseudopotential 
calculations.

Currently, dynamical correlations are included at the level of the 2B 
approximation for the self-energy, 
although the scaling of the computational cost remains 
identical using a statically screened 
electron-electron interaction for the exchange and second-order 
diagrams~\cite{C60paper2018}.
Self-energy approximations like GW or T-matrix 
would restore the cubic KBE scaling with the 
number of time steps unless a GKBA for $W$ or $T$ is provided, an 
advance which would be of utmost theoretical and computational value.

So far CHEERS simulations involving photoionization 
processes have been 
performed using the KS continuum states generated either by the Quantum 
Espresso code~\cite{QuantumEspresso} (planewave basis) or by the  
Octopus code~\cite{OCTOPUS} (grid basis). Another promising 
alternative consists in using a B-spline 
basis~\cite{Ruberti2014,Ruberti2018}. In a recent work, the B-spline basis 
has been combined with the algebraic diagrammatic construction method 
to calculate the attosecond pump-probe spectrum of carbon 
dioxide~\cite{RubertiDeclevaVitali2018}.

Another limitation of CHEERS is that the nuclear positions are kept 
fixed during the time evolution. Work to include harmonic effects 
through a Fan self-energy~\cite{Fan-PhysRev.82.900} evaluated with ab-initio frequencies and 
electron-nuclear couplings is in progress. We are also planning 
to implement the semiclassical Ehrenfest dynamics which requires to 
calculate all one- and two-electron integrals along the nuclear 
trajectory. This extension of CHEERS is especially relevant to access 
the subpicosecond timescale (10$\div$100~fs) typical of charge 
transfer and charge separation processes.

\section*{Acknowledgements}
We acknowledge inspiring and insightul discussions with 
Emil B\"ostrom,
Fabio Covito,
Daniel Karlsson,
Simone Latini,
Andrea Marini,
Yaroslav Pavlyukh,
Angel Rubio,
Davide Sangalli,
Anna-Maija Uimonen,
Robert van Leeuwen
and
Claudio Verdozzi.
We also acknowledge EC funding through the RISE Co-ExAN (Grant No. GA644076).
E.P. also acknowledges funding from the European Union project 
MaX Materials design at the eXascale H2020-EINFRA-2015-1, Grant Agreement No.
676598 and Nanoscience Foundries and
Fine Analysis-Europe H2020-INFRAIA-2014-2015, Grant Agreement No. 654360.
G.S. also 
acknowledge Tor Vergata University for financial support through the Mission Sustainability Project
2DUTOPI.


\begin{thebibliography}{79}
\expandafter\ifx\csname natexlab\endcsname\relax\def\natexlab#1{#1}\fi
\expandafter\ifx\csname bibnamefont\endcsname\relax
  \def\bibnamefont#1{#1}\fi
\expandafter\ifx\csname bibfnamefont\endcsname\relax
  \def\bibfnamefont#1{#1}\fi
\expandafter\ifx\csname citenamefont\endcsname\relax
  \def\citenamefont#1{#1}\fi
\expandafter\ifx\csname url\endcsname\relax
  \def\url#1{\texttt{#1}}\fi
\expandafter\ifx\csname urlprefix\endcsname\relax\def\urlprefix{URL }\fi
\providecommand{\bibinfo}[2]{#2}
\providecommand{\eprint}[2][]{\url{#2}}

\bibitem[{\citenamefont{Krausz and Ivanov}(2009)}]{RevModPhys.81.163}
\bibinfo{author}{\bibfnamefont{F.}~\bibnamefont{Krausz}} \bibnamefont{and}
  \bibinfo{author}{\bibfnamefont{M.}~\bibnamefont{Ivanov}},
  \bibinfo{journal}{Rev. Mod. Phys.} \textbf{\bibinfo{volume}{81}},
  \bibinfo{pages}{163} (\bibinfo{year}{2009}),
  \urlprefix\url{https://link.aps.org/doi/10.1103/RevModPhys.81.163}.

\bibitem[{\citenamefont{Gallmann et~al.}(2012)\citenamefont{Gallmann, Cirelli,
  and Keller}}]{Gallmann-review}
\bibinfo{author}{\bibfnamefont{L.}~\bibnamefont{Gallmann}},
  \bibinfo{author}{\bibfnamefont{C.}~\bibnamefont{Cirelli}}, \bibnamefont{and}
  \bibinfo{author}{\bibfnamefont{U.}~\bibnamefont{Keller}},
  \bibinfo{journal}{Annual Review of Physical Chemistry}
  \textbf{\bibinfo{volume}{63}}, \bibinfo{pages}{447} (\bibinfo{year}{2012}),
  \urlprefix\url{https://doi.org/10.1146/annurev-physchem-032511-143702}.

\bibitem[{\citenamefont{Nisoli et~al.}(2017)\citenamefont{Nisoli, Decleva,
  Calegari, Palacios, and Mart\'in}}]{Nisoli-review}
\bibinfo{author}{\bibfnamefont{M.}~\bibnamefont{Nisoli}},
  \bibinfo{author}{\bibfnamefont{P.}~\bibnamefont{Decleva}},
  \bibinfo{author}{\bibfnamefont{F.}~\bibnamefont{Calegari}},
  \bibinfo{author}{\bibfnamefont{A.}~\bibnamefont{Palacios}}, \bibnamefont{and}
  \bibinfo{author}{\bibfnamefont{F.}~\bibnamefont{Mart\'in}},
  \bibinfo{journal}{Chemical Reviews} \textbf{\bibinfo{volume}{117}},
  \bibinfo{pages}{10760} (\bibinfo{year}{2017}), \bibinfo{note}{pMID:
  28488433}, \urlprefix\url{http://dx.doi.org/10.1021/acs.chemrev.6b00453}.

\bibitem[{\citenamefont{Danielewicz}(1984)}]{danielewicz1984quantum}
\bibinfo{author}{\bibfnamefont{P.}~\bibnamefont{Danielewicz}},
  \bibinfo{journal}{Annals of Physics} \textbf{\bibinfo{volume}{152}},
  \bibinfo{pages}{239} (\bibinfo{year}{1984}).

\bibitem[{\citenamefont{Stefanucci and van Leeuwen}(2013)}]{svl-book}
\bibinfo{author}{\bibfnamefont{G.}~\bibnamefont{Stefanucci}} \bibnamefont{and}
  \bibinfo{author}{\bibfnamefont{R.}~\bibnamefont{van Leeuwen}},
  \emph{\bibinfo{title}{Nonequilibrium Many-Body Theory of Quantum Systems: A
  Modern Introduction}} (\bibinfo{publisher}{Cambridge University Press},
  \bibinfo{address}{Cambridge}, \bibinfo{year}{2013}).

\bibitem[{\citenamefont{Balzer and Bonitz}(2012)}]{balzer2012nonequilibrium}
\bibinfo{author}{\bibfnamefont{K.}~\bibnamefont{Balzer}} \bibnamefont{and}
  \bibinfo{author}{\bibfnamefont{M.}~\bibnamefont{Bonitz}},
  \emph{\bibinfo{title}{Nonequilibrium Green's Functions Approach to
  Inhomogeneous Systems}} (\bibinfo{publisher}{Springer},
  \bibinfo{year}{2012}).

\bibitem[{\citenamefont{Konstantinov and
  Perel}(1961)}]{konstantinov1961diagram}
\bibinfo{author}{\bibfnamefont{O.}~\bibnamefont{Konstantinov}}
  \bibnamefont{and} \bibinfo{author}{\bibfnamefont{V.}~\bibnamefont{Perel}},
  \bibinfo{journal}{SOVIET PHYSICS JETP-USSR} \textbf{\bibinfo{volume}{12}},
  \bibinfo{pages}{142} (\bibinfo{year}{1961}).

\bibitem[{\citenamefont{Kadanoff and Baym}(1962)}]{kadanoff1962quantum}
\bibinfo{author}{\bibfnamefont{L.~P.} \bibnamefont{Kadanoff}} \bibnamefont{and}
  \bibinfo{author}{\bibfnamefont{G.~A.} \bibnamefont{Baym}},
  \emph{\bibinfo{title}{Quantum statistical mechanics: Green's function methods
  in equilibrium and nonequilibirum problems}} (\bibinfo{publisher}{Benjamin},
  \bibinfo{year}{1962}).

\bibitem[{\citenamefont{Keldysh et~al.}(1965)}]{keldysh1965diagram}
\bibinfo{author}{\bibfnamefont{L.~V.} \bibnamefont{Keldysh}}
  \bibnamefont{et~al.}, \bibinfo{journal}{Sov. Phys. JETP}
  \textbf{\bibinfo{volume}{20}}, \bibinfo{pages}{1018} (\bibinfo{year}{1965}).

\bibitem[{\citenamefont{Kwong and Bonitz}(2000)}]{PhysRevLett.84.1768}
\bibinfo{author}{\bibfnamefont{N.-H.} \bibnamefont{Kwong}} \bibnamefont{and}
  \bibinfo{author}{\bibfnamefont{M.}~\bibnamefont{Bonitz}},
  \bibinfo{journal}{Phys. Rev. Lett.} \textbf{\bibinfo{volume}{84}},
  \bibinfo{pages}{1768} (\bibinfo{year}{2000}),
  \urlprefix\url{https://link.aps.org/doi/10.1103/PhysRevLett.84.1768}.

\bibitem[{\citenamefont{Dahlen and van Leeuwen}(2007)}]{PhysRevLett.98.153004}
\bibinfo{author}{\bibfnamefont{N.~E.} \bibnamefont{Dahlen}} \bibnamefont{and}
  \bibinfo{author}{\bibfnamefont{R.}~\bibnamefont{van Leeuwen}},
  \bibinfo{journal}{Phys. Rev. Lett.} \textbf{\bibinfo{volume}{98}},
  \bibinfo{pages}{153004} (\bibinfo{year}{2007}),
  \urlprefix\url{https://link.aps.org/doi/10.1103/PhysRevLett.98.153004}.

\bibitem[{\citenamefont{Balzer et~al.}(2009)\citenamefont{Balzer, Bonitz, van
  Leeuwen, Stan, and Dahlen}}]{PhysRevB.79.245306}
\bibinfo{author}{\bibfnamefont{K.}~\bibnamefont{Balzer}},
  \bibinfo{author}{\bibfnamefont{M.}~\bibnamefont{Bonitz}},
  \bibinfo{author}{\bibfnamefont{R.}~\bibnamefont{van Leeuwen}},
  \bibinfo{author}{\bibfnamefont{A.}~\bibnamefont{Stan}}, \bibnamefont{and}
  \bibinfo{author}{\bibfnamefont{N.~E.} \bibnamefont{Dahlen}},
  \bibinfo{journal}{Phys. Rev. B} \textbf{\bibinfo{volume}{79}},
  \bibinfo{pages}{245306} (\bibinfo{year}{2009}),
  \urlprefix\url{https://link.aps.org/doi/10.1103/PhysRevB.79.245306}.

\bibitem[{\citenamefont{My\"oh\"anen et~al.}(2008)\citenamefont{My\"oh\"anen,
  Stan, Stefanucci, and van Leeuwen}}]{mssvl.2008}
\bibinfo{author}{\bibfnamefont{P.}~\bibnamefont{My\"oh\"anen}},
  \bibinfo{author}{\bibfnamefont{A.}~\bibnamefont{Stan}},
  \bibinfo{author}{\bibfnamefont{G.}~\bibnamefont{Stefanucci}},
  \bibnamefont{and} \bibinfo{author}{\bibfnamefont{R.}~\bibnamefont{van
  Leeuwen}}, \bibinfo{journal}{EPL (Europhysics Letters)}
  \textbf{\bibinfo{volume}{84}}, \bibinfo{pages}{67001} (\bibinfo{year}{2008}),
  \urlprefix\url{http://stacks.iop.org/0295-5075/84/i=6/a=67001}.

\bibitem[{\citenamefont{My\"oh\"anen et~al.}(2009)\citenamefont{My\"oh\"anen,
  Stan, Stefanucci, and van Leeuwen}}]{mssvl.2009}
\bibinfo{author}{\bibfnamefont{P.}~\bibnamefont{My\"oh\"anen}},
  \bibinfo{author}{\bibfnamefont{A.}~\bibnamefont{Stan}},
  \bibinfo{author}{\bibfnamefont{G.}~\bibnamefont{Stefanucci}},
  \bibnamefont{and} \bibinfo{author}{\bibfnamefont{R.}~\bibnamefont{van
  Leeuwen}}, \bibinfo{journal}{Phys. Rev. B} \textbf{\bibinfo{volume}{80}},
  \bibinfo{pages}{115107} (\bibinfo{year}{2009}),
  \urlprefix\url{https://link.aps.org/doi/10.1103/PhysRevB.80.115107}.

\bibitem[{\citenamefont{von Friesen et~al.}(2009)\citenamefont{von Friesen,
  Verdozzi, and Almbladh}}]{pva.2009}
\bibinfo{author}{\bibfnamefont{M.~P.} \bibnamefont{von Friesen}},
  \bibinfo{author}{\bibfnamefont{C.}~\bibnamefont{Verdozzi}}, \bibnamefont{and}
  \bibinfo{author}{\bibfnamefont{C.-O.} \bibnamefont{Almbladh}},
  \bibinfo{journal}{Phys. Rev. Lett.} \textbf{\bibinfo{volume}{103}},
  \bibinfo{pages}{176404} (\bibinfo{year}{2009}),
  \urlprefix\url{https://link.aps.org/doi/10.1103/PhysRevLett.103.176404}.

\bibitem[{\citenamefont{Puig~von Friesen et~al.}(2010)\citenamefont{Puig~von
  Friesen, Verdozzi, and Almbladh}}]{pva.2010}
\bibinfo{author}{\bibfnamefont{M.}~\bibnamefont{Puig~von Friesen}},
  \bibinfo{author}{\bibfnamefont{C.}~\bibnamefont{Verdozzi}}, \bibnamefont{and}
  \bibinfo{author}{\bibfnamefont{C.-O.} \bibnamefont{Almbladh}},
  \bibinfo{journal}{Phys. Rev. B} \textbf{\bibinfo{volume}{82}},
  \bibinfo{pages}{155108} (\bibinfo{year}{2010}),
  \urlprefix\url{https://link.aps.org/doi/10.1103/PhysRevB.82.155108}.

\bibitem[{\citenamefont{Balzer et~al.}(2010{\natexlab{a}})\citenamefont{Balzer,
  Bauch, and Bonitz}}]{PhysRevA.81.022510}
\bibinfo{author}{\bibfnamefont{K.}~\bibnamefont{Balzer}},
  \bibinfo{author}{\bibfnamefont{S.}~\bibnamefont{Bauch}}, \bibnamefont{and}
  \bibinfo{author}{\bibfnamefont{M.}~\bibnamefont{Bonitz}},
  \bibinfo{journal}{Phys. Rev. A} \textbf{\bibinfo{volume}{81}},
  \bibinfo{pages}{022510} (\bibinfo{year}{2010}{\natexlab{a}}),
  \urlprefix\url{https://link.aps.org/doi/10.1103/PhysRevA.81.022510}.

\bibitem[{\citenamefont{Balzer et~al.}(2010{\natexlab{b}})\citenamefont{Balzer,
  Bauch, and Bonitz}}]{PhysRevA.82.033427}
\bibinfo{author}{\bibfnamefont{K.}~\bibnamefont{Balzer}},
  \bibinfo{author}{\bibfnamefont{S.}~\bibnamefont{Bauch}}, \bibnamefont{and}
  \bibinfo{author}{\bibfnamefont{M.}~\bibnamefont{Bonitz}},
  \bibinfo{journal}{Phys. Rev. A} \textbf{\bibinfo{volume}{82}},
  \bibinfo{pages}{033427} (\bibinfo{year}{2010}{\natexlab{b}}),
  \urlprefix\url{https://link.aps.org/doi/10.1103/PhysRevA.82.033427}.

\bibitem[{\citenamefont{Sch\"uler et~al.}(2016)\citenamefont{Sch\"uler,
  Berakdar, and Pavlyukh}}]{PhysRevB.93.054303}
\bibinfo{author}{\bibfnamefont{M.}~\bibnamefont{Sch\"uler}},
  \bibinfo{author}{\bibfnamefont{J.}~\bibnamefont{Berakdar}}, \bibnamefont{and}
  \bibinfo{author}{\bibfnamefont{Y.}~\bibnamefont{Pavlyukh}},
  \bibinfo{journal}{Phys. Rev. B} \textbf{\bibinfo{volume}{93}},
  \bibinfo{pages}{054303} (\bibinfo{year}{2016}),
  \urlprefix\url{https://link.aps.org/doi/10.1103/PhysRevB.93.054303}.

\bibitem[{\citenamefont{Lipavsk\'y et~al.}(1986)\citenamefont{Lipavsk\'y,
  \ifmmode \check{S}\else \v{S}\fi{}pi\ifmmode~\check{c}\else \v{c}\fi{}ka, and
  Velick\'y}}]{PhysRevB.34.6933}
\bibinfo{author}{\bibfnamefont{P.}~\bibnamefont{Lipavsk\'y}},
  \bibinfo{author}{\bibfnamefont{V.}~\bibnamefont{\ifmmode \check{S}\else
  \v{S}\fi{}pi\ifmmode~\check{c}\else \v{c}\fi{}ka}}, \bibnamefont{and}
  \bibinfo{author}{\bibfnamefont{B.}~\bibnamefont{Velick\'y}},
  \bibinfo{journal}{Phys. Rev. B} \textbf{\bibinfo{volume}{34}},
  \bibinfo{pages}{6933} (\bibinfo{year}{1986}),
  \urlprefix\url{https://link.aps.org/doi/10.1103/PhysRevB.34.6933}.

\bibitem[{\citenamefont{Latini et~al.}(2014)\citenamefont{Latini, Perfetto,
  Uimonen, van Leeuwen, and Stefanucci}}]{LPUvLS.2014}
\bibinfo{author}{\bibfnamefont{S.}~\bibnamefont{Latini}},
  \bibinfo{author}{\bibfnamefont{E.}~\bibnamefont{Perfetto}},
  \bibinfo{author}{\bibfnamefont{A.-M.} \bibnamefont{Uimonen}},
  \bibinfo{author}{\bibfnamefont{R.}~\bibnamefont{van Leeuwen}},
  \bibnamefont{and}
  \bibinfo{author}{\bibfnamefont{G.}~\bibnamefont{Stefanucci}},
  \bibinfo{journal}{Phys. Rev. B} \textbf{\bibinfo{volume}{89}},
  \bibinfo{pages}{075306} (\bibinfo{year}{2014}),
  \urlprefix\url{https://link.aps.org/doi/10.1103/PhysRevB.89.075306}.

\bibitem[{\citenamefont{Pernal et~al.}(2007)\citenamefont{Pernal, Gritsenko,
  and Baerends}}]{PhysRevA.75.012506}
\bibinfo{author}{\bibfnamefont{K.}~\bibnamefont{Pernal}},
  \bibinfo{author}{\bibfnamefont{O.}~\bibnamefont{Gritsenko}},
  \bibnamefont{and} \bibinfo{author}{\bibfnamefont{E.~J.}
  \bibnamefont{Baerends}}, \bibinfo{journal}{Phys. Rev. A}
  \textbf{\bibinfo{volume}{75}}, \bibinfo{pages}{012506}
  (\bibinfo{year}{2007}),
  \urlprefix\url{https://link.aps.org/doi/10.1103/PhysRevA.75.012506}.

\bibitem[{\citenamefont{Giesbertz et~al.}(2010)\citenamefont{Giesbertz,
  Gritsenko, and Baerends}}]{Giesbertz2010}
\bibinfo{author}{\bibfnamefont{K.~J.~H.} \bibnamefont{Giesbertz}},
  \bibinfo{author}{\bibfnamefont{O.~V.} \bibnamefont{Gritsenko}},
  \bibnamefont{and} \bibinfo{author}{\bibfnamefont{E.~J.}
  \bibnamefont{Baerends}}, \bibinfo{journal}{The Journal of Chemical Physics}
  \textbf{\bibinfo{volume}{133}}, \bibinfo{pages}{174119}
  (\bibinfo{year}{2010}), \urlprefix\url{https://doi.org/10.1063/1.3499601}.

\bibitem[{\citenamefont{Giesbertz et~al.}(2014)\citenamefont{Giesbertz,
  Gritsenko, and Baerends}}]{Giesbertz2014}
\bibinfo{author}{\bibfnamefont{K.~J.~H.} \bibnamefont{Giesbertz}},
  \bibinfo{author}{\bibfnamefont{O.~V.} \bibnamefont{Gritsenko}},
  \bibnamefont{and} \bibinfo{author}{\bibfnamefont{E.~J.}
  \bibnamefont{Baerends}}, \bibinfo{journal}{The Journal of Chemical Physics}
  \textbf{\bibinfo{volume}{140}}, \bibinfo{pages}{18A517}
  (\bibinfo{year}{2014}), \urlprefix\url{https://doi.org/10.1063/1.4867000}.

\bibitem[{\citenamefont{Brics et~al.}(2016)\citenamefont{Brics, Rapp, and
  Bauer}}]{Brics2016}
\bibinfo{author}{\bibfnamefont{M.}~\bibnamefont{Brics}},
  \bibinfo{author}{\bibfnamefont{J.}~\bibnamefont{Rapp}}, \bibnamefont{and}
  \bibinfo{author}{\bibfnamefont{D.}~\bibnamefont{Bauer}},
  \bibinfo{journal}{Phys. Rev. A} \textbf{\bibinfo{volume}{93}},
  \bibinfo{pages}{013404} (\bibinfo{year}{2016}),
  \urlprefix\url{https://link.aps.org/doi/10.1103/PhysRevA.93.013404}.

\bibitem[{\citenamefont{Brics et~al.}(2017)\citenamefont{Brics, Rapp, and
  Bauer}}]{Brics2017}
\bibinfo{author}{\bibfnamefont{M.}~\bibnamefont{Brics}},
  \bibinfo{author}{\bibfnamefont{J.}~\bibnamefont{Rapp}}, \bibnamefont{and}
  \bibinfo{author}{\bibfnamefont{D.}~\bibnamefont{Bauer}},
  \bibinfo{journal}{Journal of Physics B: Atomic, Molecular and Optical
  Physics} \textbf{\bibinfo{volume}{50}}, \bibinfo{pages}{144003}
  (\bibinfo{year}{2017}),
  \urlprefix\url{http://stacks.iop.org/0953-4075/50/i=14/a=144003}.

\bibitem[{\citenamefont{Lackner
  et~al.}(2015{\natexlab{a}})\citenamefont{Lackner, B\ifmmode~\check{r}\else
  \v{r}\fi{}ezinov\'a, Sato, Ishikawa, and Burgd\"orfer}}]{Lackner-PRA2015}
\bibinfo{author}{\bibfnamefont{F.}~\bibnamefont{Lackner}},
  \bibinfo{author}{\bibfnamefont{I.}~\bibnamefont{B\ifmmode~\check{r}\else
  \v{r}\fi{}ezinov\'a}},
  \bibinfo{author}{\bibfnamefont{T.}~\bibnamefont{Sato}},
  \bibinfo{author}{\bibfnamefont{K.~L.} \bibnamefont{Ishikawa}},
  \bibnamefont{and}
  \bibinfo{author}{\bibfnamefont{J.}~\bibnamefont{Burgd\"orfer}},
  \bibinfo{journal}{Phys. Rev. A} \textbf{\bibinfo{volume}{91}},
  \bibinfo{pages}{023412} (\bibinfo{year}{2015}{\natexlab{a}}),
  \urlprefix\url{https://link.aps.org/doi/10.1103/PhysRevA.91.023412}.

\bibitem[{\citenamefont{Lackner
  et~al.}(2015{\natexlab{b}})\citenamefont{Lackner, B?ezinov‡, Sato, Ishikawa,
  and Burgdšrfer}}]{Lackner-JOPCS2015}
\bibinfo{author}{\bibfnamefont{F.}~\bibnamefont{Lackner}},
  \bibinfo{author}{\bibfnamefont{I.}~\bibnamefont{B?ezinov‡}},
  \bibinfo{author}{\bibfnamefont{T.}~\bibnamefont{Sato}},
  \bibinfo{author}{\bibfnamefont{K.~L.} \bibnamefont{Ishikawa}},
  \bibnamefont{and}
  \bibinfo{author}{\bibfnamefont{J.}~\bibnamefont{Burgdšrfer}},
  \bibinfo{journal}{Journal of Physics: Conference Series}
  \textbf{\bibinfo{volume}{635}}, \bibinfo{pages}{112084}
  (\bibinfo{year}{2015}{\natexlab{b}}),
  \urlprefix\url{http://stacks.iop.org/1742-6596/635/i=11/a=112084}.

\bibitem[{\citenamefont{Hermanns et~al.}(2014)\citenamefont{Hermanns,
  Schl\"unzen, and Bonitz}}]{HermannsPRB2014}
\bibinfo{author}{\bibfnamefont{S.}~\bibnamefont{Hermanns}},
  \bibinfo{author}{\bibfnamefont{N.}~\bibnamefont{Schl\"unzen}},
  \bibnamefont{and} \bibinfo{author}{\bibfnamefont{M.}~\bibnamefont{Bonitz}},
  \bibinfo{journal}{Phys. Rev. B} \textbf{\bibinfo{volume}{90}},
  \bibinfo{pages}{125111} (\bibinfo{year}{2014}),
  \urlprefix\url{https://link.aps.org/doi/10.1103/PhysRevB.90.125111}.

\bibitem[{\citenamefont{Schl\"unzen and Bonitz}(2016)}]{CTPPBonitz2016}
\bibinfo{author}{\bibfnamefont{N.}~\bibnamefont{Schl\"unzen}} \bibnamefont{and}
  \bibinfo{author}{\bibfnamefont{M.}~\bibnamefont{Bonitz}},
  \bibinfo{journal}{Contrib. Plasma Phys.} \textbf{\bibinfo{volume}{56}},
  \bibinfo{pages}{5} (\bibinfo{year}{2016}),
  \urlprefix\url{http://dx.doi.org/10.1002/ctpp.201610003}.

\bibitem[{\citenamefont{Bar~Lev and Reichman}(2014)}]{BarLevReichman}
\bibinfo{author}{\bibfnamefont{Y.}~\bibnamefont{Bar~Lev}} \bibnamefont{and}
  \bibinfo{author}{\bibfnamefont{D.~R.} \bibnamefont{Reichman}},
  \bibinfo{journal}{Phys. Rev. B} \textbf{\bibinfo{volume}{89}},
  \bibinfo{pages}{220201} (\bibinfo{year}{2014}),
  \urlprefix\url{https://link.aps.org/doi/10.1103/PhysRevB.89.220201}.

\bibitem[{\citenamefont{Pal et~al.}(2011)\citenamefont{Pal, Pavlyukh,
  H{\"u}bner, and Schneider}}]{Pal2011}
\bibinfo{author}{\bibfnamefont{G.}~\bibnamefont{Pal}},
  \bibinfo{author}{\bibfnamefont{Y.}~\bibnamefont{Pavlyukh}},
  \bibinfo{author}{\bibfnamefont{W.}~\bibnamefont{H{\"u}bner}},
  \bibnamefont{and} \bibinfo{author}{\bibfnamefont{H.~C.}
  \bibnamefont{Schneider}}, \bibinfo{journal}{The European Physical Journal B}
  \textbf{\bibinfo{volume}{79}}, \bibinfo{pages}{327} (\bibinfo{year}{2011}),
  \urlprefix\url{https://doi.org/10.1140/epjb/e2010-10033-1}.

\bibitem[{\citenamefont{Perfetto
  et~al.}(2015{\natexlab{a}})\citenamefont{Perfetto, Sangalli, Marini, and
  Stefanucci}}]{PSMS.2015}
\bibinfo{author}{\bibfnamefont{E.}~\bibnamefont{Perfetto}},
  \bibinfo{author}{\bibfnamefont{D.}~\bibnamefont{Sangalli}},
  \bibinfo{author}{\bibfnamefont{A.}~\bibnamefont{Marini}}, \bibnamefont{and}
  \bibinfo{author}{\bibfnamefont{G.}~\bibnamefont{Stefanucci}},
  \bibinfo{journal}{Phys. Rev. B} \textbf{\bibinfo{volume}{92}},
  \bibinfo{pages}{205304} (\bibinfo{year}{2015}{\natexlab{a}}),
  \urlprefix\url{https://link.aps.org/doi/10.1103/PhysRevB.92.205304}.

\bibitem[{\citenamefont{Sangalli et~al.}(2016)\citenamefont{Sangalli,
  Dal~Conte, Manzoni, Cerullo, and Marini}}]{Sangalli-2016}
\bibinfo{author}{\bibfnamefont{D.}~\bibnamefont{Sangalli}},
  \bibinfo{author}{\bibfnamefont{S.}~\bibnamefont{Dal~Conte}},
  \bibinfo{author}{\bibfnamefont{C.}~\bibnamefont{Manzoni}},
  \bibinfo{author}{\bibfnamefont{G.}~\bibnamefont{Cerullo}}, \bibnamefont{and}
  \bibinfo{author}{\bibfnamefont{A.}~\bibnamefont{Marini}},
  \bibinfo{journal}{Phys. Rev. B} \textbf{\bibinfo{volume}{93}},
  \bibinfo{pages}{195205} (\bibinfo{year}{2016}),
  \urlprefix\url{https://link.aps.org/doi/10.1103/PhysRevB.93.195205}.

\bibitem[{\citenamefont{Pogna et~al.}(2016)\citenamefont{Pogna, Marsili,
  De~Fazio, Dal~Conte, Manzoni, Sangalli, Yoon, Lombardo, Ferrari, Marini
  et~al.}}]{Pogna.2016}
\bibinfo{author}{\bibfnamefont{E.~A.~A.} \bibnamefont{Pogna}},
  \bibinfo{author}{\bibfnamefont{M.}~\bibnamefont{Marsili}},
  \bibinfo{author}{\bibfnamefont{D.}~\bibnamefont{De~Fazio}},
  \bibinfo{author}{\bibfnamefont{S.}~\bibnamefont{Dal~Conte}},
  \bibinfo{author}{\bibfnamefont{C.}~\bibnamefont{Manzoni}},
  \bibinfo{author}{\bibfnamefont{D.}~\bibnamefont{Sangalli}},
  \bibinfo{author}{\bibfnamefont{D.}~\bibnamefont{Yoon}},
  \bibinfo{author}{\bibfnamefont{A.}~\bibnamefont{Lombardo}},
  \bibinfo{author}{\bibfnamefont{A.~C.} \bibnamefont{Ferrari}},
  \bibinfo{author}{\bibfnamefont{A.}~\bibnamefont{Marini}},
  \bibnamefont{et~al.}, \bibinfo{journal}{ACS Nano}
  \textbf{\bibinfo{volume}{10}}, \bibinfo{pages}{1182} (\bibinfo{year}{2016}),
  \urlprefix\url{https://doi.org/10.1021/acsnano.5b06488}.

\bibitem[{\citenamefont{{Sangalli, D.} and {Marini,
  A.}}(2015)}]{SangalliEPL2015}
\bibinfo{author}{\bibnamefont{{Sangalli, D.}}} \bibnamefont{and}
  \bibinfo{author}{\bibnamefont{{Marini, A.}}}, \bibinfo{journal}{EPL}
  \textbf{\bibinfo{volume}{110}}, \bibinfo{pages}{47004}
  (\bibinfo{year}{2015}),
  \urlprefix\url{https://doi.org/10.1209/0295-5075/110/47004}.

\bibitem[{\citenamefont{Perfetto
  et~al.}(2016{\natexlab{a}})\citenamefont{Perfetto, Sangalli, Marini, and
  Stefanucci}}]{PSMS.2016}
\bibinfo{author}{\bibfnamefont{E.}~\bibnamefont{Perfetto}},
  \bibinfo{author}{\bibfnamefont{D.}~\bibnamefont{Sangalli}},
  \bibinfo{author}{\bibfnamefont{A.}~\bibnamefont{Marini}}, \bibnamefont{and}
  \bibinfo{author}{\bibfnamefont{G.}~\bibnamefont{Stefanucci}},
  \bibinfo{journal}{Phys. Rev. B} \textbf{\bibinfo{volume}{94}},
  \bibinfo{pages}{245303} (\bibinfo{year}{2016}{\natexlab{a}}),
  \urlprefix\url{https://link.aps.org/doi/10.1103/PhysRevB.94.245303}.

\bibitem[{\citenamefont{Perfetto
  et~al.}(2015{\natexlab{b}})\citenamefont{Perfetto, Uimonen, van Leeuwen, and
  Stefanucci}}]{PUvLS.2015}
\bibinfo{author}{\bibfnamefont{E.}~\bibnamefont{Perfetto}},
  \bibinfo{author}{\bibfnamefont{A.-M.} \bibnamefont{Uimonen}},
  \bibinfo{author}{\bibfnamefont{R.}~\bibnamefont{van Leeuwen}},
  \bibnamefont{and}
  \bibinfo{author}{\bibfnamefont{G.}~\bibnamefont{Stefanucci}},
  \bibinfo{journal}{Phys. Rev. A} \textbf{\bibinfo{volume}{92}},
  \bibinfo{pages}{033419} (\bibinfo{year}{2015}{\natexlab{b}}),
  \urlprefix\url{https://link.aps.org/doi/10.1103/PhysRevA.92.033419}.

\bibitem[{\citenamefont{Bostr\"om et~al.}(2018)\citenamefont{Bostr\"om,
  Mikkelsen, Verdozzi, Perfetto, and Stefanucci}}]{C60paper2018}
\bibinfo{author}{\bibfnamefont{E.~V.} \bibnamefont{Bostr\"om}},
  \bibinfo{author}{\bibfnamefont{A.}~\bibnamefont{Mikkelsen}},
  \bibinfo{author}{\bibfnamefont{C.}~\bibnamefont{Verdozzi}},
  \bibinfo{author}{\bibfnamefont{E.}~\bibnamefont{Perfetto}}, \bibnamefont{and}
  \bibinfo{author}{\bibfnamefont{G.}~\bibnamefont{Stefanucci}},
  \bibinfo{journal}{Nano Lett.} \textbf{\bibinfo{volume}{18}},
  \bibinfo{pages}{785} (\bibinfo{year}{2018}),
  \urlprefix\url{https://doi.org/10.1021/acs.nanolett.7b03995}.

\bibitem[{\citenamefont{Perfetto et~al.}(2018)\citenamefont{Perfetto, Sangalli,
  Marini, and Stefanucci}}]{PSMS.2018}
\bibinfo{author}{\bibfnamefont{E.}~\bibnamefont{Perfetto}},
  \bibinfo{author}{\bibfnamefont{D.}~\bibnamefont{Sangalli}},
  \bibinfo{author}{\bibfnamefont{A.}~\bibnamefont{Marini}}, \bibnamefont{and}
  \bibinfo{author}{\bibfnamefont{G.}~\bibnamefont{Stefanucci}},
  \bibinfo{journal}{The Journal of Physical Chemistry Letters}
  \textbf{\bibinfo{volume}{9}}, \bibinfo{pages}{1353} (\bibinfo{year}{2018}),
  \urlprefix\url{https://doi.org/10.1021/acs.jpclett.8b00025}.

\bibitem[{\citenamefont{Covito et~al.}(2018{\natexlab{a}})\citenamefont{Covito,
  Perfetto, Rubio, and Stefanucci}}]{Covito2018}
\bibinfo{author}{\bibfnamefont{F.}~\bibnamefont{Covito}},
  \bibinfo{author}{\bibfnamefont{E.}~\bibnamefont{Perfetto}},
  \bibinfo{author}{\bibfnamefont{A.}~\bibnamefont{Rubio}}, \bibnamefont{and}
  \bibinfo{author}{\bibfnamefont{G.}~\bibnamefont{Stefanucci}},
  \bibinfo{journal}{Phys. Rev. A} \textbf{\bibinfo{volume}{97}},
  \bibinfo{pages}{061401} (\bibinfo{year}{2018}{\natexlab{a}}),
  \urlprefix\url{https://link.aps.org/doi/10.1103/PhysRevA.97.061401}.

\bibitem[{\citenamefont{Luttinger}(1964)}]{Luttinger-thermal}
\bibinfo{author}{\bibfnamefont{J.~M.} \bibnamefont{Luttinger}},
  \bibinfo{journal}{Phys. Rev.} \textbf{\bibinfo{volume}{135}},
  \bibinfo{pages}{A1505} (\bibinfo{year}{1964}),
  \urlprefix\url{https://link.aps.org/doi/10.1103/PhysRev.135.A1505}.

\bibitem[{\citenamefont{Eich et~al.}(2014)\citenamefont{Eich, Principi,
  Di~Ventra, and Vignale}}]{EichPrincipiVentraVignale:14}
\bibinfo{author}{\bibfnamefont{F.~G.} \bibnamefont{Eich}},
  \bibinfo{author}{\bibfnamefont{A.}~\bibnamefont{Principi}},
  \bibinfo{author}{\bibfnamefont{M.}~\bibnamefont{Di~Ventra}},
  \bibnamefont{and} \bibinfo{author}{\bibfnamefont{G.}~\bibnamefont{Vignale}},
  \bibinfo{journal}{Phys. Rev. B} \textbf{\bibinfo{volume}{90}},
  \bibinfo{pages}{115116} (\bibinfo{year}{2014}),
  \urlprefix\url{https://link.aps.org/doi/10.1103/PhysRevB.90.115116}.

\bibitem[{\citenamefont{Eich et~al.}(2016)\citenamefont{Eich, Di~Ventra, and
  Vignale}}]{EDVV.2016}
\bibinfo{author}{\bibfnamefont{F.~G.} \bibnamefont{Eich}},
  \bibinfo{author}{\bibfnamefont{M.}~\bibnamefont{Di~Ventra}},
  \bibnamefont{and} \bibinfo{author}{\bibfnamefont{G.}~\bibnamefont{Vignale}},
  \bibinfo{journal}{Phys. Rev. B} \textbf{\bibinfo{volume}{93}},
  \bibinfo{pages}{134309} (\bibinfo{year}{2016}),
  \urlprefix\url{https://link.aps.org/doi/10.1103/PhysRevB.93.134309}.

\bibitem[{\citenamefont{Covito et~al.}(2018{\natexlab{b}})\citenamefont{Covito,
  Eich, Tuovinen, Sentef, and Rubio}}]{CETSR.2018}
\bibinfo{author}{\bibfnamefont{F.}~\bibnamefont{Covito}},
  \bibinfo{author}{\bibfnamefont{F.~G.} \bibnamefont{Eich}},
  \bibinfo{author}{\bibfnamefont{R.}~\bibnamefont{Tuovinen}},
  \bibinfo{author}{\bibfnamefont{M.~A.} \bibnamefont{Sentef}},
  \bibnamefont{and} \bibinfo{author}{\bibfnamefont{A.}~\bibnamefont{Rubio}},
  \bibinfo{journal}{Journal of Chemical Theory and Computation}
  \textbf{\bibinfo{volume}{14}}, \bibinfo{pages}{2495}
  (\bibinfo{year}{2018}{\natexlab{b}}),
  \urlprefix\url{https://doi.org/10.1021/acs.jctc.8b00077}.

\bibitem[{\citenamefont{Galperin}(2017)}]{GalperinCSR2017}
\bibinfo{author}{\bibfnamefont{M.}~\bibnamefont{Galperin}},
  \bibinfo{journal}{Chem. Soc. Rev.} \textbf{\bibinfo{volume}{46}},
  \bibinfo{pages}{4000} (\bibinfo{year}{2017}),
  \urlprefix\url{http://dx.doi.org/10.1039/C7CS00067G}.

\bibitem[{\citenamefont{Meir and Wingreen}(1992)}]{MW92}
\bibinfo{author}{\bibfnamefont{Y.}~\bibnamefont{Meir}} \bibnamefont{and}
  \bibinfo{author}{\bibfnamefont{N.~S.} \bibnamefont{Wingreen}},
  \bibinfo{journal}{Phys. Rev. Lett.} \textbf{\bibinfo{volume}{68}},
  \bibinfo{pages}{2512} (\bibinfo{year}{1992}),
  \urlprefix\url{https://link.aps.org/doi/10.1103/PhysRevLett.68.2512}.

\bibitem[{\citenamefont{Jauho et~al.}(1994)\citenamefont{Jauho, Wingreen, and
  Meir}}]{JWM94}
\bibinfo{author}{\bibfnamefont{A.-P.} \bibnamefont{Jauho}},
  \bibinfo{author}{\bibfnamefont{N.~S.} \bibnamefont{Wingreen}},
  \bibnamefont{and} \bibinfo{author}{\bibfnamefont{Y.}~\bibnamefont{Meir}},
  \bibinfo{journal}{Phys. Rev. B} \textbf{\bibinfo{volume}{50}},
  \bibinfo{pages}{5528} (\bibinfo{year}{1994}),
  \urlprefix\url{https://link.aps.org/doi/10.1103/PhysRevB.50.5528}.

\bibitem[{\citenamefont{Sch\"uler and Pavlyukh}(2018)}]{schuler2017}
\bibinfo{author}{\bibfnamefont{M.}~\bibnamefont{Sch\"uler}} \bibnamefont{and}
  \bibinfo{author}{\bibfnamefont{Y.}~\bibnamefont{Pavlyukh}},
  \bibinfo{journal}{Phys. Rev. B} \textbf{\bibinfo{volume}{97}},
  \bibinfo{pages}{115164} (\bibinfo{year}{2018}),
  \urlprefix\url{https://link.aps.org/doi/10.1103/PhysRevB.97.115164}.

\bibitem[{\citenamefont{Dahlen and van Leeuwen}(2005)}]{DahlenLeeuwen2005}
\bibinfo{author}{\bibfnamefont{N.~E.} \bibnamefont{Dahlen}} \bibnamefont{and}
  \bibinfo{author}{\bibfnamefont{R.}~\bibnamefont{van Leeuwen}},
  \bibinfo{journal}{The Journal of Chemical Physics}
  \textbf{\bibinfo{volume}{122}}, \bibinfo{pages}{164102}
  (\bibinfo{year}{2005}), \urlprefix\url{https://doi.org/10.1063/1.1884965}.

\bibitem[{\citenamefont{Balzer et~al.}(2012)\citenamefont{Balzer, Hermanns, and
  Bonitz}}]{BalzerHermanns2012}
\bibinfo{author}{\bibfnamefont{K.}~\bibnamefont{Balzer}},
  \bibinfo{author}{\bibfnamefont{S.}~\bibnamefont{Hermanns}}, \bibnamefont{and}
  \bibinfo{author}{\bibfnamefont{M.}~\bibnamefont{Bonitz}},
  \bibinfo{journal}{EPL (Europhysics Letters)} \textbf{\bibinfo{volume}{98}},
  \bibinfo{pages}{67002} (\bibinfo{year}{2012}),
  \urlprefix\url{http://stacks.iop.org/0295-5075/98/i=6/a=67002}.

\bibitem[{\citenamefont{S\"akkinen et~al.}(2012)\citenamefont{S\"akkinen,
  Manninen, and van Leeuwen}}]{Sakkinen-2012}
\bibinfo{author}{\bibfnamefont{N.}~\bibnamefont{S\"akkinen}},
  \bibinfo{author}{\bibfnamefont{M.}~\bibnamefont{Manninen}}, \bibnamefont{and}
  \bibinfo{author}{\bibfnamefont{R.}~\bibnamefont{van Leeuwen}},
  \bibinfo{journal}{New Journal of Physics} \textbf{\bibinfo{volume}{14}},
  \bibinfo{pages}{013032} (\bibinfo{year}{2012}),
  \urlprefix\url{http://stacks.iop.org/1367-2630/14/i=1/a=013032}.

\bibitem[{\citenamefont{Hopjan et~al.}(2016)\citenamefont{Hopjan, Karlsson,
  Ydman, Verdozzi, and Almbladh}}]{HopjanPRL2016}
\bibinfo{author}{\bibfnamefont{M.}~\bibnamefont{Hopjan}},
  \bibinfo{author}{\bibfnamefont{D.}~\bibnamefont{Karlsson}},
  \bibinfo{author}{\bibfnamefont{S.}~\bibnamefont{Ydman}},
  \bibinfo{author}{\bibfnamefont{C.}~\bibnamefont{Verdozzi}}, \bibnamefont{and}
  \bibinfo{author}{\bibfnamefont{C.-O.} \bibnamefont{Almbladh}},
  \bibinfo{journal}{Phys. Rev. Lett.} \textbf{\bibinfo{volume}{116}},
  \bibinfo{pages}{236402} (\bibinfo{year}{2016}),
  \urlprefix\url{https://link.aps.org/doi/10.1103/PhysRevLett.116.236402}.

\bibitem[{\citenamefont{Lev and Reichman}(2016)}]{ReichmanEPL2016}
\bibinfo{author}{\bibfnamefont{Y.~B.} \bibnamefont{Lev}} \bibnamefont{and}
  \bibinfo{author}{\bibfnamefont{D.~R.} \bibnamefont{Reichman}},
  \bibinfo{journal}{EPL (Europhysics Letters)} \textbf{\bibinfo{volume}{113}},
  \bibinfo{pages}{46001} (\bibinfo{year}{2016}),
  \urlprefix\url{http://stacks.iop.org/0295-5075/113/i=4/a=46001}.

\bibitem[{\citenamefont{Schl\"unzen et~al.}(2017)\citenamefont{Schl\"unzen,
  Joost, Heidrich-Meisner, and Bonitz}}]{Joost2017}
\bibinfo{author}{\bibfnamefont{N.}~\bibnamefont{Schl\"unzen}},
  \bibinfo{author}{\bibfnamefont{J.-P.} \bibnamefont{Joost}},
  \bibinfo{author}{\bibfnamefont{F.}~\bibnamefont{Heidrich-Meisner}},
  \bibnamefont{and} \bibinfo{author}{\bibfnamefont{M.}~\bibnamefont{Bonitz}},
  \bibinfo{journal}{Phys. Rev. B} \textbf{\bibinfo{volume}{95}},
  \bibinfo{pages}{165139} (\bibinfo{year}{2017}),
  \urlprefix\url{https://link.aps.org/doi/10.1103/PhysRevB.95.165139}.

\bibitem[{\citenamefont{{A.-M.~Uimonen}
  et~al.}(2011)\citenamefont{{A.-M.~Uimonen}, Khosravi, Stan, Stefanucci,
  Kurth, {R.~van~Leeuwen}, and {E.K.U.~Gross}}}]{UKSSKvLG.2011}
\bibinfo{author}{\bibnamefont{{A.-M.~Uimonen}}},
  \bibinfo{author}{\bibfnamefont{E.}~\bibnamefont{Khosravi}},
  \bibinfo{author}{\bibfnamefont{A.}~\bibnamefont{Stan}},
  \bibinfo{author}{\bibfnamefont{G.}~\bibnamefont{Stefanucci}},
  \bibinfo{author}{\bibfnamefont{S.}~\bibnamefont{Kurth}},
  \bibinfo{author}{\bibnamefont{{R.~van~Leeuwen}}}, \bibnamefont{and}
  \bibinfo{author}{\bibnamefont{{E.K.U.~Gross}}}, \bibinfo{journal}{Phys. Rev.
  B} \textbf{\bibinfo{volume}{84}}, \bibinfo{pages}{115103}
  (\bibinfo{year}{2011}).

\bibitem[{\citenamefont{Almbladh et~al.}(1989)\citenamefont{Almbladh, Morales,
  and Grossmann}}]{PhysRevB.39.3489}
\bibinfo{author}{\bibfnamefont{C.-O.} \bibnamefont{Almbladh}},
  \bibinfo{author}{\bibfnamefont{A.~L.} \bibnamefont{Morales}},
  \bibnamefont{and}
  \bibinfo{author}{\bibfnamefont{G.}~\bibnamefont{Grossmann}},
  \bibinfo{journal}{Phys. Rev. B} \textbf{\bibinfo{volume}{39}},
  \bibinfo{pages}{3489} (\bibinfo{year}{1989}),
  \urlprefix\url{https://link.aps.org/doi/10.1103/PhysRevB.39.3489}.

\bibitem[{\citenamefont{Almbladh and Morales}(1989)}]{PhysRevB.39.3503}
\bibinfo{author}{\bibfnamefont{C.-O.} \bibnamefont{Almbladh}} \bibnamefont{and}
  \bibinfo{author}{\bibfnamefont{A.~L.} \bibnamefont{Morales}},
  \bibinfo{journal}{Phys. Rev. B} \textbf{\bibinfo{volume}{39}},
  \bibinfo{pages}{3503} (\bibinfo{year}{1989}),
  \urlprefix\url{https://link.aps.org/doi/10.1103/PhysRevB.39.3503}.

\bibitem[{\citenamefont{Cini}(1977)}]{cini1993two}
\bibinfo{author}{\bibfnamefont{M.}~\bibnamefont{Cini}}, \bibinfo{journal}{Solid
  State Communications} \textbf{\bibinfo{volume}{24}}, \bibinfo{pages}{681 }
  (\bibinfo{year}{1977}),
  \urlprefix\url{http://www.sciencedirect.com/science/article/pii/0038109877903908}.

\bibitem[{\citenamefont{Sawatzky}(1977)}]{PhysRevLett.39.504}
\bibinfo{author}{\bibfnamefont{G.~A.} \bibnamefont{Sawatzky}},
  \bibinfo{journal}{Phys. Rev. Lett.} \textbf{\bibinfo{volume}{39}},
  \bibinfo{pages}{504} (\bibinfo{year}{1977}),
  \urlprefix\url{https://link.aps.org/doi/10.1103/PhysRevLett.39.504}.

\bibitem[{\citenamefont{Stefanucci et~al.}(2008)\citenamefont{Stefanucci,
  Kurth, Rubio, and Gross}}]{Stefanuccipumping}
\bibinfo{author}{\bibfnamefont{G.}~\bibnamefont{Stefanucci}},
  \bibinfo{author}{\bibfnamefont{S.}~\bibnamefont{Kurth}},
  \bibinfo{author}{\bibfnamefont{A.}~\bibnamefont{Rubio}}, \bibnamefont{and}
  \bibinfo{author}{\bibfnamefont{E.~K.~U.} \bibnamefont{Gross}},
  \bibinfo{journal}{Phys. Rev. B} \textbf{\bibinfo{volume}{77}},
  \bibinfo{pages}{075339} (\bibinfo{year}{2008}),
  \urlprefix\url{https://link.aps.org/doi/10.1103/PhysRevB.77.075339}.

\bibitem[{\citenamefont{Stefanucci et~al.}(2010)\citenamefont{Stefanucci,
  Perfetto, and Cini}}]{spc.2010}
\bibinfo{author}{\bibfnamefont{G.}~\bibnamefont{Stefanucci}},
  \bibinfo{author}{\bibfnamefont{E.}~\bibnamefont{Perfetto}}, \bibnamefont{and}
  \bibinfo{author}{\bibfnamefont{M.}~\bibnamefont{Cini}},
  \bibinfo{journal}{Phys. Rev. B} \textbf{\bibinfo{volume}{81}},
  \bibinfo{pages}{115446} (\bibinfo{year}{2010}),
  \urlprefix\url{https://link.aps.org/doi/10.1103/PhysRevB.81.115446}.

\bibitem[{\citenamefont{Kalvov\'a et~al.}(2018)\citenamefont{Kalvov\'a,
  Velick\'y, and \ifmmode \check{S}\else \v{S}\fi{}pi\ifmmode~\check{c}\else
  \v{c}\fi{}ka}}]{Kalvova.2017}
\bibinfo{author}{\bibfnamefont{A.}~\bibnamefont{Kalvov\'a}},
  \bibinfo{author}{\bibfnamefont{B.}~\bibnamefont{Velick\'y}},
  \bibnamefont{and} \bibinfo{author}{\bibfnamefont{V.}~\bibnamefont{\ifmmode
  \check{S}\else \v{S}\fi{}pi\ifmmode~\check{c}\else \v{c}\fi{}ka}},
  \bibinfo{journal}{EPL (Europhysics Letters)} \textbf{\bibinfo{volume}{121}},
  \bibinfo{pages}{67002} (\bibinfo{year}{2018}),
  \urlprefix\url{http://stacks.iop.org/0295-5075/121/i=6/a=67002}.

\bibitem[{\citenamefont{Goulielmakis et~al.}(2010)\citenamefont{Goulielmakis,
  Loh, Wirth, Santra, Rohringer, Yakovlev, Zherebtsov, Pfeifer, Azzeer, Kling
  et~al.}}]{goulielmakis2010real}
\bibinfo{author}{\bibfnamefont{E.}~\bibnamefont{Goulielmakis}},
  \bibinfo{author}{\bibfnamefont{Z.-H.} \bibnamefont{Loh}},
  \bibinfo{author}{\bibfnamefont{A.}~\bibnamefont{Wirth}},
  \bibinfo{author}{\bibfnamefont{R.}~\bibnamefont{Santra}},
  \bibinfo{author}{\bibfnamefont{N.}~\bibnamefont{Rohringer}},
  \bibinfo{author}{\bibfnamefont{V.~S.} \bibnamefont{Yakovlev}},
  \bibinfo{author}{\bibfnamefont{S.}~\bibnamefont{Zherebtsov}},
  \bibinfo{author}{\bibfnamefont{T.}~\bibnamefont{Pfeifer}},
  \bibinfo{author}{\bibfnamefont{A.~M.} \bibnamefont{Azzeer}},
  \bibinfo{author}{\bibfnamefont{M.~F.} \bibnamefont{Kling}},
  \bibnamefont{et~al.}, \bibinfo{journal}{Nature}
  \textbf{\bibinfo{volume}{466}}, \bibinfo{pages}{739} (\bibinfo{year}{2010}).

\bibitem[{\citenamefont{Bunge et~al.}(1993)\citenamefont{Bunge, Barrientos, and
  Bunge}}]{BUNGE1993113}
\bibinfo{author}{\bibfnamefont{C.}~\bibnamefont{Bunge}},
  \bibinfo{author}{\bibfnamefont{J.}~\bibnamefont{Barrientos}},
  \bibnamefont{and} \bibinfo{author}{\bibfnamefont{A.}~\bibnamefont{Bunge}},
  \bibinfo{journal}{Atomic Data and Nuclear Data Tables}
  \textbf{\bibinfo{volume}{53}}, \bibinfo{pages}{113 } (\bibinfo{year}{1993}),
  ISSN \bibinfo{issn}{0092-640X},
  \urlprefix\url{http://www.sciencedirect.com/science/article/pii/S0092640X8371003X}.

\bibitem[{smi()}]{smiles1}
\bibinfo{note}{J. Fern\'andez Rico, I. Ema, R. L\'opez, G. Ram\'irez and K.
  Ishida, in {\em Recent Advances in Computational Chemistry: Molecular
  Integrals over Slater Orbitals}, eds. T. Ozdogan and M. B. Ruiz (Transworld
  Research Network, 2008), pp. 145.}

\bibitem[{\citenamefont{Rico et~al.}(2004)\citenamefont{Rico, L\'opez, Ema, and
  Ram'rez}}]{smiles2}
\bibinfo{author}{\bibfnamefont{J.~F.} \bibnamefont{Rico}},
  \bibinfo{author}{\bibfnamefont{R.}~\bibnamefont{L\'opez}},
  \bibinfo{author}{\bibfnamefont{I.}~\bibnamefont{Ema}}, \bibnamefont{and}
  \bibinfo{author}{\bibfnamefont{G.}~\bibnamefont{Ram'rez}},
  \bibinfo{journal}{Journal of Computational Chemistry}
  \textbf{\bibinfo{volume}{25}}, \bibinfo{pages}{1987} (\bibinfo{year}{2004}),
  \urlprefix\url{https://onlinelibrary.wiley.com/doi/abs/10.1002/jcc.20131}.

\bibitem[{\citenamefont{Giannozzi et~al.}(2009)\citenamefont{Giannozzi, Baroni,
  Bonini, Calandra, Car, Cavazzoni, Ceresoli, Chiarotti, Cococcioni, Dabo
  et~al.}}]{QuantumEspresso}
\bibinfo{author}{\bibfnamefont{P.}~\bibnamefont{Giannozzi}},
  \bibinfo{author}{\bibfnamefont{S.}~\bibnamefont{Baroni}},
  \bibinfo{author}{\bibfnamefont{N.}~\bibnamefont{Bonini}},
  \bibinfo{author}{\bibfnamefont{M.}~\bibnamefont{Calandra}},
  \bibinfo{author}{\bibfnamefont{R.}~\bibnamefont{Car}},
  \bibinfo{author}{\bibfnamefont{C.}~\bibnamefont{Cavazzoni}},
  \bibinfo{author}{\bibfnamefont{D.}~\bibnamefont{Ceresoli}},
  \bibinfo{author}{\bibfnamefont{G.~L.} \bibnamefont{Chiarotti}},
  \bibinfo{author}{\bibfnamefont{M.}~\bibnamefont{Cococcioni}},
  \bibinfo{author}{\bibfnamefont{I.}~\bibnamefont{Dabo}}, \bibnamefont{et~al.},
  \bibinfo{journal}{Journal of Physics: Condensed Matter}
  \textbf{\bibinfo{volume}{21}}, \bibinfo{pages}{395502}
  (\bibinfo{year}{2009}),
  \urlprefix\url{http://stacks.iop.org/0953-8984/21/i=39/a=395502}.

\bibitem[{\citenamefont{Troullier and Martins}(1991)}]{PhysRevB.43.1993}
\bibinfo{author}{\bibfnamefont{N.}~\bibnamefont{Troullier}} \bibnamefont{and}
  \bibinfo{author}{\bibfnamefont{J.~L.} \bibnamefont{Martins}},
  \bibinfo{journal}{Phys. Rev. B} \textbf{\bibinfo{volume}{43}},
  \bibinfo{pages}{1993} (\bibinfo{year}{1991}),
  \urlprefix\url{https://link.aps.org/doi/10.1103/PhysRevB.43.1993}.

\bibitem[{\citenamefont{Perdew et~al.}(1996)\citenamefont{Perdew, Burke, and
  Ernzerhof}}]{PhysRevLett.77.3865}
\bibinfo{author}{\bibfnamefont{J.~P.} \bibnamefont{Perdew}},
  \bibinfo{author}{\bibfnamefont{K.}~\bibnamefont{Burke}}, \bibnamefont{and}
  \bibinfo{author}{\bibfnamefont{M.}~\bibnamefont{Ernzerhof}},
  \bibinfo{journal}{Phys. Rev. Lett.} \textbf{\bibinfo{volume}{77}},
  \bibinfo{pages}{3865} (\bibinfo{year}{1996}),
  \urlprefix\url{https://link.aps.org/doi/10.1103/PhysRevLett.77.3865}.

\bibitem[{\citenamefont{Marini et~al.}(2009)\citenamefont{Marini, Hogan,
  Gr\"uning, and Varsano}}]{MARINI20091392}
\bibinfo{author}{\bibfnamefont{A.}~\bibnamefont{Marini}},
  \bibinfo{author}{\bibfnamefont{C.}~\bibnamefont{Hogan}},
  \bibinfo{author}{\bibfnamefont{M.}~\bibnamefont{Gr\"uning}},
  \bibnamefont{and} \bibinfo{author}{\bibfnamefont{D.}~\bibnamefont{Varsano}},
  \bibinfo{journal}{Computer Physics Communications}
  \textbf{\bibinfo{volume}{180}}, \bibinfo{pages}{1392 }
  (\bibinfo{year}{2009}),
  \urlprefix\url{http://www.sciencedirect.com/science/article/pii/S0010465509000472}.

\bibitem[{\citenamefont{Clementi and Roetti}(1974)}]{ClementiRoetti}
\bibinfo{author}{\bibfnamefont{E.}~\bibnamefont{Clementi}} \bibnamefont{and}
  \bibinfo{author}{\bibfnamefont{C.}~\bibnamefont{Roetti}},
  \bibinfo{journal}{Atomic Data and Nuclear Data Tables}
  \textbf{\bibinfo{volume}{14}}, \bibinfo{pages}{177 } (\bibinfo{year}{1974}),
  \urlprefix\url{http://www.sciencedirect.com/science/article/pii/S0092640X74800161}.

\bibitem[{\citenamefont{Andrade et~al.}(2015)\citenamefont{Andrade, Strubbe,
  De~Giovannini, Larsen, Oliveira, Alberdi-Rodriguez, Varas, Theophilou,
  Helbig, Verstraete et~al.}}]{OCTOPUS}
\bibinfo{author}{\bibfnamefont{X.}~\bibnamefont{Andrade}},
  \bibinfo{author}{\bibfnamefont{D.}~\bibnamefont{Strubbe}},
  \bibinfo{author}{\bibfnamefont{U.}~\bibnamefont{De~Giovannini}},
  \bibinfo{author}{\bibfnamefont{A.~H.} \bibnamefont{Larsen}},
  \bibinfo{author}{\bibfnamefont{M.~J.~T.} \bibnamefont{Oliveira}},
  \bibinfo{author}{\bibfnamefont{J.}~\bibnamefont{Alberdi-Rodriguez}},
  \bibinfo{author}{\bibfnamefont{A.}~\bibnamefont{Varas}},
  \bibinfo{author}{\bibfnamefont{I.}~\bibnamefont{Theophilou}},
  \bibinfo{author}{\bibfnamefont{N.}~\bibnamefont{Helbig}},
  \bibinfo{author}{\bibfnamefont{M.~J.} \bibnamefont{Verstraete}},
  \bibnamefont{et~al.}, \bibinfo{journal}{Phys. Chem. Chem. Phys.}
  \textbf{\bibinfo{volume}{17}}, \bibinfo{pages}{31371} (\bibinfo{year}{2015}),
  \urlprefix\url{http://dx.doi.org/10.1039/C5CP00351B}.

\bibitem[{\citenamefont{Perdew and Zunger}(1981)}]{PZ81}
\bibinfo{author}{\bibfnamefont{J.~P.} \bibnamefont{Perdew}} \bibnamefont{and}
  \bibinfo{author}{\bibfnamefont{A.}~\bibnamefont{Zunger}},
  \bibinfo{journal}{Phys. Rev. B} \textbf{\bibinfo{volume}{23}},
  \bibinfo{pages}{5048} (\bibinfo{year}{1981}),
  \urlprefix\url{https://link.aps.org/doi/10.1103/PhysRevB.23.5048}.

\bibitem[{\citenamefont{Perfetto
  et~al.}(2016{\natexlab{b}})\citenamefont{Perfetto, Uimonen, van Leeuwen, and
  Stefanucci}}]{PUvLS.2016}
\bibinfo{author}{\bibfnamefont{E.}~\bibnamefont{Perfetto}},
  \bibinfo{author}{\bibfnamefont{A.-M.} \bibnamefont{Uimonen}},
  \bibinfo{author}{\bibfnamefont{R.}~\bibnamefont{van Leeuwen}},
  \bibnamefont{and}
  \bibinfo{author}{\bibfnamefont{G.}~\bibnamefont{Stefanucci}},
  \bibinfo{journal}{Journal of Physics: Conference Series}
  \textbf{\bibinfo{volume}{696}}, \bibinfo{pages}{012004}
  (\bibinfo{year}{2016}{\natexlab{b}}),
  \urlprefix\url{http://stacks.iop.org/1742-6596/696/i=1/a=012004}.

\bibitem[{\citenamefont{Ruberti et~al.}(2014)\citenamefont{Ruberti, Averbukh,
  and Decleva}}]{Ruberti2014}
\bibinfo{author}{\bibfnamefont{M.}~\bibnamefont{Ruberti}},
  \bibinfo{author}{\bibfnamefont{V.}~\bibnamefont{Averbukh}}, \bibnamefont{and}
  \bibinfo{author}{\bibfnamefont{P.}~\bibnamefont{Decleva}},
  \bibinfo{journal}{The Journal of Chemical Physics}
  \textbf{\bibinfo{volume}{141}}, \bibinfo{pages}{164126}
  (\bibinfo{year}{2014}), \urlprefix\url{https://doi.org/10.1063/1.4900444}.

\bibitem[{\citenamefont{Ruberti et~al.}(2018)\citenamefont{Ruberti, Decleva,
  and Averbukh}}]{Ruberti2018}
\bibinfo{author}{\bibfnamefont{M.}~\bibnamefont{Ruberti}},
  \bibinfo{author}{\bibfnamefont{P.}~\bibnamefont{Decleva}}, \bibnamefont{and}
  \bibinfo{author}{\bibfnamefont{V.}~\bibnamefont{Averbukh}},
  \bibinfo{journal}{Phys. Chem. Chem. Phys.} \textbf{\bibinfo{volume}{20}},
  \bibinfo{pages}{8311} (\bibinfo{year}{2018}),
  \urlprefix\url{http://dx.doi.org/10.1039/C7CP07849H}.

\bibitem[{\citenamefont{Ruberti et~al.}(0)\citenamefont{Ruberti, Decleva, and
  Averbukh}}]{RubertiDeclevaVitali2018}
\bibinfo{author}{\bibfnamefont{M.}~\bibnamefont{Ruberti}},
  \bibinfo{author}{\bibfnamefont{P.}~\bibnamefont{Decleva}}, \bibnamefont{and}
  \bibinfo{author}{\bibfnamefont{V.}~\bibnamefont{Averbukh}},
  \bibinfo{journal}{Journal of Chemical Theory and Computation}
  \textbf{\bibinfo{volume}{0}}, \bibinfo{pages}{null} (\bibinfo{year}{0}),
  \urlprefix\url{https://doi.org/10.1021/acs.jctc.8b00479}.

\bibitem[{\citenamefont{Fan}(1951)}]{Fan-PhysRev.82.900}
\bibinfo{author}{\bibfnamefont{H.~Y.} \bibnamefont{Fan}},
  \bibinfo{journal}{Phys. Rev.} \textbf{\bibinfo{volume}{82}},
  \bibinfo{pages}{900} (\bibinfo{year}{1951}),
  \urlprefix\url{https://link.aps.org/doi/10.1103/PhysRev.82.900}.

\end{thebibliography}

\end{document}